\documentstyle[aps]{revtex}

\begin{document}
\title{Renormalization of the quantum chromodynamics with massive gluons }
\author{Jun-Chen Su and Hai-Jun Wang}
\address{Center for Theoretical Physics, School of Physics\\
Changchun 130023, People's Republic of China}
\maketitle
\date{}

\begin{abstract}
In our previously published papers, it was proved that the chromodynamics
with massive gluons can well be set up on the gauge-invariance principle.
The quantization of the chromodynamics was perfectly performed in the both
of Hamiltonian and Lagrangian path-integral formalisms by using the
Lagrangian undetermined multiplier method. In this paper, It is shown that
the quantum theory is invariant with respect to a kind of
BRST-transformations. From the BRST-invariance of the theory , the
Ward-Takahashi identities satisfied by the generating functionals of full
Green functions, connected Green functions and proper vertex functions are
successively derived. As an application of the above Ward-Takahashi
identities, the Ward-Takahashi identities obeyed by the massive gluon and
ghost particle propagators and various proper vertices are derived and based
on these identities, the propagators and vertices are perfectly
renormalized. Especially, as a result of the renormalization, the
Slavnov-Taylor identity satisfied by renormalization constants is natually
deduced. To demonstrate the renormalizability of the theory, the one-loop
renormalization of the theory is carried out by means of the mass-dependent
momentum space subtraction scheme and the renormalization group approach,
giving an exact one-loop effective coupling constant and one-loop effective
gluon and quark masses which show the asymptotically free behaviors as the
same as those given in the quantum chromodynamics with massless gluons.

\end{abstract}

\section{Introduction}

According to the conventional concept of quantum chromodynamics (QCD), in
order to keep the Lagrangian to be gauge-invariant, the gluons must be
massless. On the contrary, in many previous investigations of glueballs, an
effective gluon mass was phenomenologically introduced so as to get
reasonable theoretical results [1-4].\ The gluon mass was supposed to be
generated dynamically from the interaction with the physical vacuum of the
Yang-Mills theory [3] or through strong gluon-binding force [5]. Apparently,
these arguments would not be considered to be stringent and logically
consistent with the concept of the ordinary QCD. In our previous papers
[6-8], it was argued that the QCD with massive gluons, as a non-Abelian
massive gauge field theory in which the masses of all gauge fields are the
same, can actually be set up on the principle of gauge-invariance without
need of introducing the Higgs mechanism. The essential points to achieve
this conclusion are as follows. (a) The gluon fields must be viewed as a
constrained system in the whole space of vector potentials and the Lorentz
condition, as a necessary constraint, must be introduced from the beginning
and imposed on the massive Yang-Mills Lagrangian; (b) The gauge-invariance\
of a gauge field theory should be generally examined from its action other
than from the Lagrangian because the action is of more fundamental dynamical
meaning than the Lagrangian. Particularly, for a constrained system such as
the gluon field, the gauge-invariance should be seen from its action given
in the physical subspace defined by the Lorentz condition because the fields
exist and move only in the physical subspace; (c) In the physical subspace,
only infinitesimal gauge transformations are possibly allowed and necessary
to be considered in the examination of whether the theory is gauge-invariant
or not; This fact was clarified originally in Ref. [9]; (d) To construct a
correct gauge field theory, the residual gauge degrees of freedom existing
in the physical subspace must be eliminated by the constraint condition on
the gauge group. This constraint condition may be determined by requiring
the action to be gauge-invariant. Based on these points of view, it is easy
to prove that the QCD with massive gluons established in our previous papers
[6-8] is gauge-invariant.

In Refs. [6-8], the quantization of the QCD with massive gluons (will be
called massive QCD later on) was Lorentz-covariantly performed in the both
of Hamiltonian and Lagrangian path-integral formalisms by employing the
Lagrange undetermined multiplier method. In this paper, it will be shown
that the quantum theory has an important property that the effective action
appearing in the generating functional of Green functions is invariant with
respect to a kind of BRST-transformations [10]. Thus, the theory is set up
from beginning to end on the basis of gauge-invariance principle. From the
BRST-symmetry of the theory, we will derive various Ward-Takahashi (W-T)
identities [11-17] satisfied by the generating functionals of Green
functions and proper vertex functions. These W-T identities are of special
importance in proofs of unitarity and renormalizability of the theory [18].
Furthermore, from the W-T identities obeyed by the generating functionals,
we will derive W-T identities satisfied by the massive gluon propagator, the
massive ghost particle propagator, the gluon three-line proper vertex, the
gluon four-line proper vertex and the quark-gluon proper vertex which appear
in the perturbative expansion of S-matrix elements. Based on these W-T
identities, the aforementioned propagators and vertices will be perfectly
renormalized. As a result of the renormalizations, the Slavnov-Taylor (S-T)
identity satisfied by the renormalization constants [19, 20] will be derived
and shown to be formally the same as that given in the QCD with massless
gluons (will be called massless QCD\ hereafter). This identity is much
useful for practical calculations of the renormalization by the approach of
renormalization group equation [21-23]. It should be mentioned that in the
previous literature [16,24-28], the massive non-Abelian gauge field theory
without involving Higgs bosons in it was not considered to be renormalizable
and/or unitary. This conclusion was drawn from the theories which were not
established correctly because the unphysical degrees of freedom involved in
the theories are not eliminated at all by introducing appropriate constraint
conditions. In our theory, the unphysical degrees of freedom appearing in
the theory, i.e., the unphysical longitudinal components of vector
potentials for the gluon fields and the residual gauge degrees of freedom
existing in the subspace defined by Lorentz condition are respectively
eliminated by the introduced Lorentz condition and the ghost equation which
acts as the constraint condition on the gauge group. This guarantees that
the massive QCD established in our previous papers is not only unitary, but
also renormalizable. To demonstrate further the renormalizability of the
theory, in this paper, the one loop renormalization will specifically be
carried out by means of the mass-dependent momentum space subtraction scheme
and the renormalization group equation (RGE), giving an exact one-loop
effective coupling constant and one-loop effective quark and gluon masses
without any ambiguity.

The arrangement of this paper is as follows. In section 2, we will derive
the BRST-transformations under which the effective action of the massive QCD
is invariant. In section 3, we will derive the W-T identities satisfied by
various generating functionals. In section 4, to illustrate applications of
the above W-T identity, the W-T identities obeyed by the massive gluon
propagator and ghost propagator will be derived and the renormalization of
these propagators will be discussed. In sections 5-7, the W-T identities
obeyed by the gluon three-line vertex, the gluon four-line vertex and the
quark-gluon vertex will be derived and the renormalizations of the vertices
will be discussed, respectively. Section 8 serves to derive the one-loop
effective coupling constant and the effective gluon mass. Section 9 is used
to derive the one-loop effective quark masses. In the last section, some
conclusions and discussions are made. In Appendix, the W-T identities will
be given by an alternative derivation.

\section{BRST- transformation}

In Ref. [7], the QCD with massive gluons is set up by starting from the
Lagrangian 
\begin{equation}
{\cal L}=\bar \psi \{i\gamma ^\mu (\partial _\mu -igT^aA_\mu ^a)-m\}\psi -%
\frac 14F^{a\mu \nu }F_{\mu \nu }^a+\frac 12M^2A^{a\mu }A_\mu ^a  \eqnum{2.1}
\end{equation}
where $\psi (x)$ denotes the quark field function , $\bar \psi (x)$ is its
Dirac-conjugate, $T^a=\lambda ^a/2$ are the color matrices (the generators
of gauge group $SU(3)$), $m$ is the quark mass and $M$ is the gluon mass.
The above Lagrangian is constrained by the Lorentz condition 
\begin{equation}
\partial ^\mu A_\mu ^a=0.  \eqnum{2.2}
\end{equation}
Under this condition, as was proved in Refs. [6-8], the action given by the
Lagrangian in Eq. (2.1) is invariant with respect to the following gauge
transformations: 
\begin{equation}
\begin{array}{c}
\delta A_\mu ^a=\xi D_\mu ^{ab}C^b, \\ 
\delta \psi (x)=ig\xi T^aC^a(x)\psi (x), \\ 
\delta \bar \psi (x)=ig\xi \bar \psi (x)T^aC^a(x)
\end{array}
\eqnum{2.3}
\end{equation}
where 
\begin{equation}
D_\mu ^{ab}=\delta ^{ab}\partial _\mu -gf^{abc}A_\mu ^c  \eqnum{2.4}
\end{equation}
is the covariant derivative. In the above, we have set the parametric
functions of the gauge group $\theta ^a(x)=\xi C^a(x)$ in which $\xi $ is an
infinitesimal Grassmann number and $C^a(x)$ are the ghost field functions.

The quantization of the massive QCD was carried out by different approaches
in Ref. [6]. A simpler quantization is performed in the Lagrangian
path-integral formalism by means of the Lagrange undetermined multiplier
method \ which was shown to be equivalent to the Faddeev-Popov approach of
quantization [9]. For this quantization, it is convenient to generalize the\
QCD Lagrangian and the Lorentz condition to the following forms: 
\begin{equation}
\begin{array}{c}
{\cal L}_\lambda =\bar \psi \{i\gamma ^\mu (\partial _\mu -igT^aA_\mu
^a)-m\}\psi -\frac 14F^{a\mu \nu }F_{\mu \nu }^a \\ 
+\frac 12M^2A^{a\mu }A_\mu ^a-\frac 12\alpha (\lambda ^a)^2
\end{array}
\eqnum{2.5}
\end{equation}
and

\begin{equation}
\partial ^\mu A_\mu ^a+\alpha \lambda ^a=0  \eqnum{2.6}
\end{equation}
where $\lambda ^a(x)$ are the extra functions which will be identified with
the Lagrange multipliers and $\alpha $ is an arbitrary constant playing the
role of gauge parameter. According to the general procedure for constrained
systems, the constraint in Eq. (2.6) may be incorporated into the Lagrangian
in Eq. (2.5) by the Lagrange multiplier method, giving a generalized
Lagrangian such that 
\begin{equation}
\begin{array}{c}
{\cal L}_\lambda =\bar \psi \{i\gamma ^\mu (\partial _\mu -igT^aA_\mu
^a)-m\}\psi -\frac 14F^{a\mu \nu }F_{\mu \nu }^a+\frac 12M^2A^{a\mu }A_\mu ^a
\\ 
+\lambda ^a\partial ^\mu A_\mu ^a+\frac 12\alpha (\lambda ^a)^2.
\end{array}
\eqnum{2.7}
\end{equation}
This Lagrangian is obviously not gauge-invariant. However, for building up a
correct gauge field theory, it is necessary to require the dynamics of the
system, i.e. the action given by the Lagrangian (2.7) to be invariant under
the gauge transformations\ denoted in Eq. (2.3). By this requirement,
noticing the identity $f^{abc}A^{a\mu }A_\mu ^b=0$ and applying the
constraint condition in Eq.(2.6), we find

\begin{equation}
\delta S_\lambda =-\frac \xi \alpha \int d^4x\partial ^\nu A_\nu
^a(x)\partial ^\mu ({\cal D}_\mu ^{ab}(x)C^b(x))=0  \eqnum{2.8}
\end{equation}
where 
\begin{equation}
{\cal D}_\mu ^{ab}(x)=\delta ^{ab}\frac{\sigma ^2}{\Box _x}\partial _\mu
^x+D_\mu ^{ab}(x)  \eqnum{2.9}
\end{equation}
in which $\sigma ^2=\alpha M^2$, $\Box _x$ is the D'Alembertian operator and 
$D_\mu ^{ab}(x)$ was defined in Eq. (2.4). From Eq. (2.6) we see $\frac 1%
\alpha \partial ^\nu A_\nu ^a=-\lambda ^a\neq 0$. Therefore, to ensure the
action to be gauge-invariant, the following constraint condition on the
gauge group is necessary to be required 
\begin{equation}
\partial _x^\mu ({\cal D}_\mu ^{ab}(x)C^b(x))=0  \eqnum{2.10}
\end{equation}
which usually is called ghost equation. When this constraint condition is
incorporated into the Lagrangian in Eq. (2.7) by the Lagrange multiplier
method, we obtain a more generalized Lagrangian as follows 
\begin{equation}
\begin{array}{c}
{\cal L}_\lambda =\bar \psi \{i\gamma ^\mu (\partial _\mu -igT^aA_\mu
^a)-m\}\psi -\frac 14F^{a\mu \nu }F_{\mu \nu }^a+\frac 12M^2A^{a\mu }A_\mu ^a
\\ 
+\lambda ^a\partial ^\mu A_\mu ^a+\frac 12\alpha (\lambda ^a)^2+\bar C%
^a\partial ^\mu ({\cal D}_\mu ^{ab}C^b)
\end{array}
\eqnum{2.11}
\end{equation}
where $\bar C^a(x)$, acting as Lagrange undetermined multipliers, are the
new scalar variables conjugate to the ghost variables $C^a(x).$

As we learn from the Lagrange undetermined multiplier method, the dynamical
and constrained variables as well as the Lagrange multipliers in the
Lagrangian (2.11) can all be treated as free ones, varying arbitrarily.
Therefore, we are allowed to use this kind of Lagrangian to construct the
generating functional of Green functions 
\begin{equation}
\begin{array}{c}
Z[J^{a\mu },\overline{\eta },\eta ,\overline{\xi }^a,\xi ^a]=\frac 1N\int
D(A_\mu ^a,\bar \psi ,\psi ,\bar C^a,C^a,\lambda ^a)exp\{i\int d^4x[{\cal L}%
_\lambda (x) \\ 
+J^{a\mu }(x)A_\mu ^a(x)+\bar \psi \eta +\overline{\eta }\psi +\overline{\xi 
}^a(x)C^a(x)+\bar C^a(x)\xi ^a(x)]\} \\ 
=\frac 1N\int D(A_\mu ^a,\bar \psi ,\psi ,\bar C^a,C^a,)exp\{iS+\int
d^4x[J^{a\mu }(x)A_\mu ^a(x) \\ 
+\bar \psi \eta +\overline{\eta }\psi +\overline{\xi }^a(x)C^a(x)+\bar C%
^a(x)\xi ^a(x)]\}
\end{array}
\eqnum{2.12}
\end{equation}
where the last equality is obtained by carrying out the integral over $%
\lambda ^a(x)$, $D(A_\mu ^a,\cdots ,\lambda ^a)$ denotes the functional
integration measure, $J_\mu ^a,\overline{\eta },\eta ,\overline{\xi }^a$ and 
$\xi ^a$ are the external sources coupled to the gluon, quark and ghost
fields, $N$ is the normalization constant and 
\begin{equation}
\begin{array}{c}
S=\int d^4x[\bar \psi \{i\gamma ^\mu (\partial _\mu -igT^aA_\mu ^a)-m\}\psi -%
\frac 14F^{a\mu \nu }F_{\mu \nu }^a \\ 
+\frac 12M^2A^{a\mu }A_\mu ^a-\frac 1{2\alpha }(\partial ^\mu A_\mu
^a)^2-\partial ^\mu \bar C^a{\cal D}_\mu ^{ab}C^b]
\end{array}
\eqnum{2.13}
\end{equation}
is the effective action given in arbitrary gauges.

Similar to the massless QCD, for the massive QCD, there are a set of
BRST-transformations including the infinitesimal gauge transformations shown
in Eq. (2.3) and the transformations for the ghost fields under which the
effective action is invariant. The transformations for the ghost fields may
be found from the stationary condition of the effective action under the
BRST-transformations. By applying the transformations in Eq. (2.3) to the
action in Eq. (2.13), one can derive 
\begin{equation}
\delta S=\int d^4x\{[\delta \bar C^a-\frac \xi \alpha \partial ^\nu A_\nu
^a]\partial ^\mu ({\cal D}_\mu ^{ab}C^b)+\bar C^a\partial ^\mu \delta ({\cal %
D}_\mu ^{ab}C^b)\}=0.  \eqnum{2.14}
\end{equation}
This expression suggests that if we set 
\begin{equation}
\delta \bar C^a=\frac \xi \alpha \partial ^\nu A_\nu ^a  \eqnum{2.15}
\end{equation}
and 
\begin{equation}
\partial ^\mu \delta ({\cal D}_\mu ^{ab}C^b)=0.  \eqnum{2.16}
\end{equation}
The action will be invariant. Eq. (2.15) gives the transformation law of the
ghost field variable $\bar C^a(x)$ which is the same as the one in the
massless gauge field theory. From Eq. (2.16), we may derive a transformation
law of the ghost field variables $C^a(x)$. Noticing the relation in Eq.
(2.9), we can write 
\begin{equation}
\delta ({\cal D}_\mu ^{ab}(x)C^b(x))=\frac{\sigma ^2}{\Box _x}\partial _\mu
^x\delta C^a(x)+\delta (D_\mu ^{ab}(x)C^b(x)).  \eqnum{2.17}
\end{equation}
In the massless QCD, it has been proved that [13-17]

\begin{equation}
\delta (D_\mu ^{ab}(x)C^b(x))=D_\mu ^{ab}(x)[\delta C^b(x)+\frac \xi 2%
gf^{bcd}C^c(x)C^d(x)].  \eqnum{2.18}
\end{equation}
With this result, Eq. (2.17) can be written as 
\begin{equation}
\delta ({\cal D}_\mu ^{ab}(x)C^b(x))={\cal D}_\mu ^{ab}(x)\delta
C^b(x)-D_\mu ^{ab}(x)\delta C_0^b(x)  \eqnum{2.19}
\end{equation}
where 
\begin{equation}
\delta C_0^a(x)\equiv -\frac{\xi g}2f^{abc}C^b(x)C^c(x).  \eqnum{2.20}
\end{equation}
On substituting Eq. (2.19) into Eq. (2.16), we have 
\begin{equation}
M^{ab}(x)\delta C^b(x)=M_0^{ab}(x)\delta C_0^b(x)  \eqnum{2.21}
\end{equation}
where we have defined 
\begin{equation}
M^{ab}(x)\equiv \partial _x^\mu {\cal D}_\mu ^{ab}(x)=\delta ^{ab}(\Box
_x+\sigma ^2)-gf^{abc}A_\mu ^c(x)\partial _x^\mu  \eqnum{2.22}
\end{equation}
and

\begin{equation}
M_0^{ab}(x)\equiv \partial _x^\mu D_\mu ^{ab}(x)=M^{ab}(x)-\sigma ^2\delta
^{ab}.  \eqnum{2.23}
\end{equation}
It is noted that the operator in Eq. (2.22 ) is just the operator appearing
in Eq. (2.10). Corresponding to Eq.(2.10), we may write an equation
satisfied by the Green function $\Delta ^{ab}(x-y)$ 
\begin{equation}
M^{ac}(x)\Delta ^{cb}(x-y)=\delta ^{ab}\delta ^4(x-y).  \eqnum{2.24}
\end{equation}
The function $\Delta ^{ab}(x-y)$ is nothing but the exact propagator of the
ghost field which is the inverse of the operator $M^{ab}(x)$. In the light
of Eq. (2.24) and noticing Eq. (2.23), we may solve out the $\delta C^a(x)$
from Eq. (2.21) 
\begin{equation}
\begin{array}{c}
\delta C^a(x)=(M^{-1}M_0\delta C_0)^a(x)=\{M^{-1}(M-\sigma ^2)\delta
C_0\}^a(x) \\ 
=\delta C_0^a(x)-\sigma ^2\int d^4y\Delta ^{ab}(x-y)\delta C_0^b(y).
\end{array}
\eqnum{2.25}
\end{equation}
This just is the transformation law for the ghost field variables $C^a(x)$.
When the gluon mass $M$ tends to zero, Eq. (2.25) immediately goes over to
the corresponding transformation given in the massless gauge field theory.
It is interesting that in the Landau gauge ($\alpha =0),$ due to $\sigma =0$%
, the above transformation also reduces to the form as given in the massless
theory. This result is natural since in the Landau gauge, the gluon field
mass term in the action is gauge-invariant. However, in general gauges, the
mass term is no longer gauge-invariant. In this case, to maintain the action
to be gauge-invariant, it is necessary to give the ghost field a mass $%
\sigma $ so as to counteract the gauge-non-invariance of the gluon field
mass term. As a result, in the transformation given in Eq. (2.25) appears a
term proportional to $\sigma ^2$.

\section{Ward-Takahashi identities}

This section is devoted to deriving the W-T identities for massive QCD on
the basis of the BRST-symmetry of the theory. Since the derivations are much
similar to those for the QCD with massless gluons, we only need here to give
a brief description of the derivations. When we make the
BRST-transformations shown in Eqs. (2.3), (2.15) and (2.25) to the
generating functional in Eq. (2.12) and consider the invariance of the
generating functional, the action and the integration measure under the
transformations (the invariance of the integration measure is easy to
check), we obtain an identity such that [11-17] 
\begin{equation}
\begin{array}{c}
\frac 1N\int {\cal D}(A_\mu ^a,\bar C^a,C^a,\bar \psi ,\psi )\int
d^4x\{J^{a\mu }(x)\delta A_\mu ^a(x)+\delta \overline{C}^a(x)\xi ^a(x)+%
\overline{\xi }^a(x)\delta C^a(x) \\ 
+\bar \eta (x)\delta \psi (x)+\delta \bar \psi (x)\eta (x)\}e^{iS+EST} \\ 
=0
\end{array}
\eqnum{3.1}
\end{equation}
where $EST$ is an abbreviation of the external source terms appearing in Eq.
(2.12). The Grassmann number $\xi $ contained in the BRST-transformations in
Eq. (3.1) may be eliminated by performing a partial differentiation of Eq.
(3.1) with respect to $\xi $. As a result, we get a W-T identity as follows 
\begin{equation}
\begin{array}{c}
\frac 1N\int {\cal D}(A_\mu ^a,\bar C^a,C^a,\bar \psi ,\psi )\int
d^4x\{J^{a\mu }(x)\Delta A_\mu ^a(x)+\triangle \overline{C}^a(x)\xi ^a(x)-%
\overline{\xi }^a(x)\Delta C^a(x) \\ 
-\bar \eta (x)\Delta \psi (x)+\Delta \bar \psi (x)\eta (x)\}e^{iS+EST} \\ 
=0
\end{array}
\eqnum{3.2}
\end{equation}
where 
\begin{equation}
\begin{array}{c}
{\Delta }A_\mu ^a(x)=D_\mu ^{ab}(x)C^b(x), \\ 
{\Delta }\bar C^a(x)=\frac 1\alpha \partial ^\mu A_\mu ^a(x), \\ 
{\Delta }C^a(x)=\int d^4y[\delta ^{ab}\delta ^4(x-y)-\sigma ^2\Delta
^{ab}(x-y)]{\triangle }C_0^b(y), \\ 
{\Delta }C_0^b(y)=-\frac 12gf^{bcd}C^c(y)C^d(y), \\ 
{\Delta }\psi (x)=igT^aC^a(x)\psi (x), \\ 
{\Delta }\bar \psi (x)=ig\bar \psi (x)T^aC^a(x).
\end{array}
\eqnum{3.3}
\end{equation}
These functions defined above are finite. Each of them differs from the
corresponding BRST-transformation written in Eqs. (2.3), (2.15) and (2.25)
by an infinitesimal Grassmann parameter $\xi .$

In order to represent the composite field functions $\Delta A_\mu ^a,\Delta
C^a,\Delta \bar \psi $ and $\Delta \psi $ in Eq. (3.2) in terms of
differentials of the functional $Z$ with respect to external sources, we
may, as usual, construct a generalized generating functional by introducing
new external sources (called BRST-sources later on) into the generating
functional written in Eq. (2.12), as shown in the following [13-16] 
\begin{equation}
\begin{array}{c}
Z[J_\mu ^a,\overline{\xi }^a,\xi ^a,\bar \eta ,\eta ;u^{a\mu },v^a,\bar \zeta
,\zeta ] \\ 
=\frac 1N\int {\cal D}[A_\mu ^a,\bar C^a,C^a,\bar \psi ,\psi ]exp\{iS+i\int
d^4x[u^{a\mu }\Delta A_\mu ^a+v^a\Delta C^a \\ 
+\Delta \bar \psi \zeta +\bar \zeta \Delta \psi +J^{a\mu }A_\mu ^a+\overline{%
\xi }^aC^a+\bar C^a\xi ^a+\bar \eta \psi +\bar \psi \eta ]\}
\end{array}
\eqnum{3.4}
\end{equation}
where $u^{a\mu },$ $v^a$, $\overline{\varsigma }$ and $\varsigma $ are the
sources which belong to the corresponding functions $\Delta A_{\mu \text{ , }%
}^a\Delta C^a$, $\Delta \Psi $ and $\Delta \overline{\psi }$, respectively.
Obviously, the $u^{a\mu }$ and $\Delta A_\mu ^a$ are anticommuting
quantities, while, the $v^a$, $\bar \zeta $, $\zeta $, $\Delta C^a$, $\Delta 
\bar \psi $ and $\Delta \psi $ are commuting ones. We may start from the
above generating functional to re-derive the W-T identity. In order that the
identity thus derived is identical to that as given in Eq. (3.2), it is
necessary to require the BRST-source terms $u_i\Delta \Phi _i$, where $%
u_i=u^{a\mu }$, $v^a$, $\overline{\zeta }$ or $\zeta $ and $\Delta \Phi
_i=\Delta A_\mu ^a$, $\Delta C^a$, $\Delta \Psi $ or $\Delta \overline{\Psi }
$ to be invariant under the BRST-transformations. How to ensure the
BRST-invariance of the source terms? For illustration, let us introduce the
source terms in such a fashion 
\begin{equation}
\begin{array}{c}
\int d^4x[\widetilde{u}^{a\mu }\delta A_\mu ^a+\widetilde{v}^a\delta C^a+%
\overline{\widetilde{\zeta }}\delta \psi +\delta \overline{\psi }\widetilde{%
\zeta }] \\ 
=\int d^4x[u^{a\mu }\triangle A_\mu ^a+v^a\triangle C^a+\overline{\zeta }%
\triangle \psi +\triangle \overline{\psi }\zeta ]
\end{array}
\eqnum{3.5}
\end{equation}
where 
\begin{equation}
u^{a\mu }=\tilde u^{a\mu }\xi ,\;\;v^a=\tilde v^a\xi ,\;\;\bar \varsigma =%
\overline{\widetilde{\varsigma }}\xi ,\;\;\varsigma =-\tilde \varsigma \xi .
\eqnum{3.6}
\end{equation}
These external sources are defined by including the Grassmann number $\xi $
and hence products of them with $\xi $ vanish. This suggests that we may
generally define the sources by the following condition 
\begin{equation}
u_i\xi =0.  \eqnum{3.7}
\end{equation}
Considering that under the BRST-transformation, the variation of the
composite field functions given in the general gauges can be represented in
the form $\delta \Delta \Phi _i=\xi \tilde \Phi _i$ where $\tilde \Phi _i$
are functions without including the parameter $\xi $, clearly, the
definition in Eq. (3.7) for the sources would guarantee the BRST- invariance
of the BRST-source terms. When the BRST-transformations in Eqs. (2.3),
(2.15) and (2.25) are made to the generating functional in Eq. (3.4), due to
the definition in Eq. (3.7) for the sources, we have $u_i\delta \Delta \Phi
_i=0$ which means that the BRST-source terms give a vanishing contribution
to the identity in Eq. (3.1). Therefore, we still obtain the identity as
shown in Eq. (3.1) except that the external source terms is now extended to
include the BRST-external source terms. This fact indicates that we may
directly insert the BRST-source terms into the exponent in Eq. (3.1) without
changing the identity itself. When performing a partial differentiation of
the identity with respect to $\xi $, we obtain a W-T identity which is the
same as written in Eq. (3.2) except that the BRST-source terms are now
included in the identity. Therefore, Eq. (3.2) may be expressed as 
\begin{equation}
\begin{array}{c}
\int d^4x[J^{a\mu }(x)\frac \delta {\delta u^{a\mu }(x)}-\overline{\xi }^a(x)%
\frac \delta {\delta v^a(x)}-\bar \eta (x)\frac \delta {\delta \bar \zeta (x)%
} \\ 
+\eta (x)\frac \delta {\delta \zeta (x)}+\frac 1\alpha \xi ^a(x)\partial
_x^\mu \frac \delta {\delta J^{a\mu }(x)}]Z[J_\mu ^a,\cdots ,\zeta ] \\ 
=0.
\end{array}
\eqnum{3.8}
\end{equation}
This is the W-T identity satisfied by the generating functional of full
Green functions.

On substituting in Eq. (3.8) the relation [13-16] 
\begin{equation}
Z=e^{iW}  \eqnum{3.9}
\end{equation}
where $W$ denotes the generating functional of connected Green functions,
one may obtain a W-T identity expressed by the functional $W$ 
\begin{equation}
\begin{array}{c}
\int d^4x[J^{a\mu }(x)\frac \delta {\delta u^{a\mu }(x)}-\overline{\xi }^a(x)%
\frac \delta {\delta v^a(x)}-\bar \eta (x)\frac \delta {\delta \bar \zeta (x)%
}+\eta (x)\frac \delta {\delta \zeta (x)} \\ 
+\frac 1\alpha \xi ^a(x)\partial _x^\mu \frac \delta {\delta J^{a\mu }(x)}%
]W[J_u^a,\cdots ,\zeta ] \\ 
=0.
\end{array}
\eqnum{3.10}
\end{equation}
From this identity, one may get another W-T identity satisfied by the
generating functional $\Gamma $ of proper (one-particle-irreducible) vertex
functions. The functional $\Gamma $ is usually defined by the following
Legendre transformation [13-16]

\begin{equation}
\begin{array}{c}
\Gamma [A^{a\mu },\bar C^a,C^a,\bar \psi ,\psi ;u_\mu ^a,v^a,\bar \zeta
,\zeta ]=W[J_\mu ^a,\overline{\xi }^a,\xi ^a,\bar \eta ,\eta ;u_\mu ^a,v^a,%
\bar \zeta ,\zeta ] \\ 
-\int d^4x[J_\mu ^aA^{a\mu }+\overline{\xi }^aC^a+\bar C^a\xi ^a+\bar \eta
\psi +\bar \psi \eta ]
\end{array}
\eqnum{3.11}
\end{equation}
where $A_\mu ^a,\bar C^a,C^a,\bar \psi $ and $\psi $ are the field variables
defined by the following functional derivatives 
\begin{equation}
\begin{array}{c}
A_\mu ^a(x)=\frac{\delta W}{\delta J^{a\mu }(x)},\;\;\bar C^a(x)=-\frac{%
\delta W}{\delta \xi ^a(x)},C^a(x)=\frac{\delta W}{\delta \overline{\xi }%
^a(x)}, \\ 
\bar \psi (x)=-\frac{\delta W}{\delta \eta (x)},\;\;\psi (x)=\frac{\delta W}{%
\delta \bar \eta (x)}.
\end{array}
\eqnum{3.12}
\end{equation}
From Eq.(3.11), it is not difficult to get the inverse transformations
[13-16] 
\begin{equation}
\begin{array}{c}
J^{a\mu }(x)=-\frac{\delta \Gamma }{\delta A_\mu ^a(x)},\;\;\overline{\xi }%
^a(x)=\frac{\delta \Gamma }{\delta C^a(x)},\xi ^a(x)=-\frac{\delta \Gamma }{%
\delta \bar C^a(x)}, \\ 
\bar \eta (x)=\frac{\delta \Gamma }{\delta \psi (x)},\;\;\eta (x)=-\frac{%
\delta \Gamma }{\delta \bar \psi (x)}.
\end{array}
\eqnum{3.13}
\end{equation}
It is obvious that 
\begin{eqnarray}
\frac{\delta W}{\delta u_\mu ^a}=\frac{\delta \Gamma }{\delta u_\mu ^a},\;\;%
\frac{\delta W}{\delta v^a}=\frac{\delta \Gamma }{\delta v^a},\;\;\frac{%
\delta W}{\delta \zeta }=\frac{\delta \Gamma }{\delta \zeta },\;\;\frac{%
\delta W}{\delta \bar \zeta }=\frac{\delta \Gamma }{\delta \bar \zeta }. 
\eqnum{3.14}
\end{eqnarray}
Employing Eqs. (3.13) and (3.14), the W-T identity in Eq. (3.10) will be
written as [13-16] 
\begin{equation}
\begin{array}{c}
\int d^4x\{\frac{\delta \Gamma }{\delta A_\mu ^a(x)}\frac{\delta \Gamma }{%
\delta u^{a\mu }(x)}+\frac{\delta \Gamma }{\delta C^a(x)}\frac{\delta \Gamma 
}{\delta v^a(x)}+\frac{\delta \Gamma }{\delta \psi (x)}\frac{\delta \Gamma }{%
\delta \bar \zeta (x)} \\ 
+\frac{\delta \Gamma }{\delta \bar \psi (x)}\frac{\delta \Gamma }{\delta
\zeta (x)}+\frac 1\alpha \partial _x^\mu A_\mu ^a(x)\frac{\delta \Gamma }{%
\delta \overline{C}^a(x)}\} \\ 
=0.
\end{array}
\eqnum{3.15}
\end{equation}
This is the W-T identity satisfied by the generating functional of proper
vertex functions.

The above identity may be represented in another form with the aid of the
so-called ghost equation of motion. The ghost equation may easily be derived
by firstly making the translation transformation: $\bar C^a\rightarrow \bar C%
^a+\bar \lambda ^a$ in Eq. (2.12) where $\bar \lambda ^a$ is an arbitrary
Grassmann variable, then differentiating Eq. (2.12) with respect to the $%
\bar \lambda ^a$ and finally setting $\overline{\lambda }^a=0$. The result
is [13-16] 
\begin{equation}
\frac 1N\int D(A_\mu ^a,\bar C^a,C^a,\bar \psi ,\psi )\{\xi ^a(x)+\partial
_x^\mu ({\cal D}_\mu ^{ab}(x)C^b(x))\}e^{iS+EST}=0.  \eqnum{3.16}
\end{equation}
When we use the generating functional defined in Eq. (3.4) and notice the
relation in Eq. (2.9), the above equation may be represented as [13-16] 
\begin{equation}
\lbrack \xi ^a(x)-i\partial _x^\mu \frac \delta {\delta u^{a\mu }(x)}-i{%
\sigma }^2\frac \delta {\delta \overline{\xi }^a(x)}]Z[J_\mu ^a,\cdots
,\zeta ]=0.  \eqnum{3.17}
\end{equation}
On substituting the relation in Eq. (3.9) into the above equation, we may
write a ghost equation satisfied by the functional $W$ such that 
\begin{equation}
\xi ^a(x)+\partial _x^\mu \frac{\delta W}{\delta u^{a\mu }(x)}+{\sigma }^2%
\frac{\delta W}{\delta \overline{\xi }^a(x)}=0.  \eqnum{3.18}
\end{equation}
From this equation, the ghost equation obeyed by the functional $\Gamma $ is
easy to be derived by virtue of Eqs. (3.12) - (3.14) [13-16] 
\begin{equation}
\frac{\delta \Gamma }{\delta \bar C^a(x)}-\partial _x^\mu \frac{\delta
\Gamma }{\delta u^{a\mu }(x)}-\sigma ^2C^a(x)=0.  \eqnum{3.19}
\end{equation}
Upon applying the above equation to the last term in Eq. (3.15). the
identity in Eq. (3.15) will be rewritten as 
\begin{equation}
\begin{array}{c}
\int d^4x\{\frac{\delta \Gamma }{\delta A_\mu ^a}\frac{\delta \Gamma }{%
\delta u^{a\mu }}+\frac{\delta \Gamma }{\delta C^a}\frac{\delta \Gamma }{%
\delta v^a}+\frac{\delta \Gamma }{\delta \psi }\frac{\delta \Gamma }{\delta 
\bar \zeta }+\frac{\delta \Gamma }{\delta \bar \psi }\frac{\delta \Gamma }{%
\delta \zeta } \\ 
+M^2\partial ^\nu A_\nu ^aC^a-\frac 1\alpha \partial ^\mu \partial ^\nu
A_\nu ^a\frac{\delta \Gamma }{\delta u^{a\mu }}\} \\ 
=0.
\end{array}
\eqnum{3.20}
\end{equation}

Now, let us define a new functional $\hat \Gamma $ in such a manner 
\begin{equation}
\hat \Gamma =\Gamma +\frac 1{2\alpha }\int d^4x(\partial ^\mu A_\mu ^a)^2. 
\eqnum{3.21}
\end{equation}
From this definition, it follows that 
\begin{equation}
\frac{\delta \Gamma }{\delta A_\mu ^a}=\frac{\delta \hat \Gamma }{\delta
A_\mu ^a}+\frac 1\alpha \partial ^\mu \partial ^\nu A_\nu ^a.  \eqnum{3.22}
\end{equation}
When inserting Eq. (3.21) into Eq. (3.20) and considering the relation in
Eq. (3.22), we arrive at 
\begin{equation}
\int d^4x\{\frac{\delta \hat \Gamma }{\delta A_\mu ^a}\frac{\delta \hat 
\Gamma }{\delta u^{a\mu }}+\frac{\delta \hat \Gamma }{\delta C^a}\frac{%
\delta \hat \Gamma }{\delta v^a}+\frac{\delta \hat \Gamma }{\delta \psi }%
\frac{\delta \hat \Gamma }{\delta \bar \zeta }+\frac{\delta \hat \Gamma }{%
\delta \bar \psi }\frac{\delta \hat \Gamma }{\delta \zeta }+M^2\partial ^\nu
A_\nu ^aC^a\}=0.  \eqnum{3.23}
\end{equation}
The ghost equation represented through the functional ${\hat \Gamma }$ is of
the same form as Eq. (3.19) 
\begin{equation}
\frac{\delta \hat \Gamma }{\delta \bar C^a(x)}-\partial _x^\mu \frac{\delta 
\hat \Gamma }{\delta u^{a\mu }(x)}-{\sigma }^2C^a(x)=0.  \eqnum{3.24}
\end{equation}
In the Landau gauge, since $\sigma =0$ and ${\partial ^\nu A_\nu ^a=0}$,
Eqs. (3.23) and (3.24) respectively reduce to [13-16] 
\begin{equation}
\int d^4x\{\frac{\delta \hat \Gamma }{\delta A_\mu ^a}\frac{\delta \hat 
\Gamma }{\delta u^{a\mu }}+\frac{\delta \hat \Gamma }{\delta C^a}\frac{%
\delta \hat \Gamma }{\delta v^a}+\frac{\delta \hat \Gamma }{\delta \psi }%
\frac{\delta \hat \Gamma }{\delta \bar \zeta }+\frac{\delta \hat \Gamma }{%
\delta \bar \psi }\frac{\delta \hat \Gamma }{\delta \zeta }\}=0  \eqnum{3.25}
\end{equation}
and 
\begin{equation}
\frac{\delta \hat \Gamma }{\delta \bar C^a}-\partial ^\mu \frac{\delta \hat 
\Gamma }{\delta u^{a\mu }}=0.  \eqnum{3.26}
\end{equation}
These equations formally are the same as those for the massless QCD.

From the W-T identities formulated above, we may derive various W-T
identities obeyed by Green functions and vertices, as will be illustrated in
the next sections.

\section{Gluon and ghost particle Propagators}

In this section, we plan to derive the W-T identities satisfied by the
massive gluon and ghost particle propagators by starting from the W-T
identity represented in Eq. (3.8) and the ghost equation shown in Eq. (3.17)
and then discuss their renormalization. Let us perform differentiations of
the identities in Eqs. (3.8) and (3.17) with respect to the external sources 
$\xi ^a(x)$ and $\xi ^b(y)$ respectively and then set all the sources except
for the source $J_\mu ^a(x)$ to be zero. In this way, we obtain the
following identities 
\begin{eqnarray}
\frac 1\alpha \partial _x^\mu \frac{\delta Z[J]}{\delta J^{a\mu }(x)}+\int
d^4yJ^{b\nu }(y)\frac{\delta ^2Z[J,\xi ,u]}{\delta \xi ^a(x)\delta u^{b\nu
}(y)}|_{\xi =u=0}=0  \eqnum{4.1}
\end{eqnarray}
and 
\begin{equation}
\begin{array}{c}
i\partial _\mu ^x\frac{\delta ^2Z[J.\xi .u]}{\delta u_\mu ^a(x)\delta \xi
^b(y)}|_{\xi =u=0}+i{\sigma }^2\frac{\delta ^2Z[J,\overline{\xi },\xi ]}{%
\delta \overline{\xi }^a(x)\delta \xi ^b(y)}|_{\overline{\xi }=\xi =0} \\ 
+\delta ^{ab}\delta ^4(x-y)Z[J]=0.
\end{array}
\eqnum{4.2}
\end{equation}
Furthermore, on differentiating Eq. (4.1) with respect to $J_\nu ^b(y)$ and
then letting the source $J$ vanish, we may get an identity which is, in
operator representation, of the form [13-16] $\ $%
\begin{equation}
\frac 1\alpha \partial _x^\mu <0^{+}|T[\hat A_\mu ^a(x)\hat A_\nu
^b(y)]|0^{-}>=<0^{+}|T^{*}[\hat {\bar C^a}(x)\hat D_\nu ^{bd}(y)\hat C%
^d(y)]|0^{-}>  \eqnum{4.3}
\end{equation}
where $\hat A_\nu ^a(x)$, $\hat C^a(x)$ and $\hat {\bar C^a}(x)$ stand for
the gluon field and ghost field operators and $T^{*}$ symbolizes the
covariant time-ordering product. When the source $J$ is set to vanish, Eq.
(4.2) gives such an equation [13-16] 
\begin{equation}
\begin{array}{c}
i\partial _y^\nu <0^{+}|T^{*}\{\hat {\bar C^a}(x)\hat D_\nu ^{bd}(y)\hat C%
^d(y)\}|0^{-}> \\ 
+i{\sigma }^2<0^{+}|T[\hat {\bar C^a}(x)\hat C^b(y)]|0^{-}>=\delta
^{ab}\delta ^4(x-y).
\end{array}
\eqnum{4.4}
\end{equation}
Upon inserting Eq. (4.3) into Eq. (4.4), we have 
\begin{equation}
\partial _x^\mu \partial _y^\nu D_{\mu \nu }^{ab}(x-y)-\alpha \sigma
^2\Delta ^{ab}(x-y)=-\alpha \delta ^{ab}\delta ^4(x-y)  \eqnum{4.5}
\end{equation}
where 
\begin{equation}
iD_{\mu \nu }^{ab}(x-y)=<0^{+}|T\{\hat A_\mu ^a(x)\hat A_\nu ^b(y)\}|0^{-}> 
\eqnum{4.6}
\end{equation}
which is the familiar full gluon propagator and 
\begin{equation}
i\Delta ^{ab}(x-y)=<0^{+}|T\{\hat C^a(x)\hat {\bar C^b}(y)\}|0^{-}> 
\eqnum{4.7}
\end{equation}
which is the full ghost particle propagator. Eq. (4.5) just is the W-T
identity respected by the gluon propagator which establishes a relation
between the longitudinal part of gluon propagator and the ghost particle
propagator. Particularly, in the Landau gauge ($\alpha =0$), as we see, Eq.
(4.5) reduces to the form which exhibits the transversity of the gluon
propagator. By the Fourier transformation, Eq. (4.5) will be converted to
the form given in the momentum space as follows 
\begin{equation}
k^\mu k^\nu D_{\mu \nu }^{ab}(k)-\alpha \sigma ^2\Delta ^{ab}(k)=-\alpha
\delta ^{ab}.  \eqnum{4.8}
\end{equation}
The ghost particle propagator may be determined by the ghost equation shown
in Eq. (4.4). However, we would rather here to derive its expression from
the Dyson-Schwinger equation [29] satisfied by the propagator which may be
established by the perturbation method. 
\begin{equation}
\Delta ^{ab}(k)=\Delta _0^{ab}(k)+\Delta _0^{aa^{\prime }}(k){\Omega }%
^{a^{\prime }b^{\prime }}(k)\Delta ^{b^{\prime }b}(k)  \eqnum{4.9}
\end{equation}
where 
\begin{equation}
i\Delta _0^{ab}(k)=i\delta ^{ab}\Delta _0(k)=\frac{-i\delta ^{ab}}{%
k^2-\sigma ^2+i\varepsilon }  \eqnum{4.10}
\end{equation}
is the free ghost particle propagator which can be derived from the
generating functional in Eq. (2.12) by a perturbative calculation and $%
-i\Omega ^{ab}(k)=-i\delta ^{ab}\Omega (k)$ denotes the proper self-energy
operator of ghost particle. From Eq. (4.9), it is easy to solve that 
\begin{equation}
i\Delta ^{ab}(k)=\frac{-i\delta ^{ab}}{k^2[1+\hat \Omega (k^2)]-\sigma
^2+i\varepsilon }  \eqnum{4.11}
\end{equation}
where the self-energy has properly been expressed as 
\begin{equation}
\Omega (k)=k^2\hat \Omega (k^2).  \eqnum{4.12}
\end{equation}
Similarly, we may write a Dyson-Schwinger equation for the gluon propagator
by the perturbation procedure$\ $[29] 
\begin{equation}
D_{\mu \nu }(k)=D_{\mu \nu }^0(k)+D_{\mu \lambda }^0(k)\Pi ^{\lambda \rho
}(k)D_{\rho \nu }(k)  \eqnum{4.13}
\end{equation}
where the color indices are suppressed for simplicity and 
\begin{equation}
iD_{\mu \nu }^{(0)ab}(k)=i\delta ^{ab}D_{\mu \nu }^{(0)}(k)=-i\delta ^{ab}[%
\frac{g_{\mu \nu }-k_\mu k_\nu /k^2}{k^2-M^2+i\varepsilon }+\frac{\alpha
k_\mu k_\nu /k^2}{k^2-\sigma ^2+i\varepsilon }]  \eqnum{4.14}
\end{equation}
is the free gluon propagator which can easily be derived from the
perturbative expansion of the generating functional in Eq. (2.12) and $-i\Pi
_{\mu \nu }^{ab}(k)=-i\delta ^{ab}\Pi _{\mu \nu }(k)$ stands for the gluon
proper self-energy operator. Let us decompose the propagator and the
self-energy operator into a transverse part and a longitudinal part: 
\begin{equation}
D^{\mu \nu }(k)=D_T^{\mu \nu }(k)+D_L^{\mu \nu }(k),\Pi ^{\mu \nu }(k)=\Pi
_T^{\mu \nu }(k)+\Pi _L^{\mu \nu }(k)  \eqnum{4.15}
\end{equation}
where 
\begin{equation}
\begin{array}{c}
D_T^{\mu \nu }(k)={\cal P}_T^{\mu \nu }(k)D_T(k^2),\text{ }D_L^{\mu \nu }(k)=%
{\cal P}_L^{\mu \nu }(k)D_L(k^2), \\ 
\Pi _T^{\mu \nu }(k)={\cal P}_T^{\mu \nu }(k)\Pi _T(k^2),\text{ }\Pi _L^{\mu
\nu }(k)={\cal P}_L^{\mu \nu }(k)\Pi _L(k^2)
\end{array}
\eqnum{4.16}
\end{equation}
here ${\cal P}_T^{\mu \nu }(k)=(g^{\mu \nu }-\frac{k^\mu k^\nu }{k^2})$ and $%
{\cal P}_L^{\mu \nu }(k)=\frac{k^\mu k^\nu }{k^2}$ are the transverse and
longitudinal projectors respectively. Considering these decompositions and
the orthogonality between the transverse and longitudinal parts, Eq. (4.13)
will be split into two equations 
\begin{equation}
D_{T\mu \nu }(k)=D_{T\mu \nu }^0(k)+D_{T\mu \lambda }^0(k)\Pi _T^{\lambda
\rho }(k)D_{T\rho \nu }(k)  \eqnum{4.17}
\end{equation}
and 
\begin{equation}
D_{L\mu \nu }(k)=D_{L\mu \nu }^0(k)+D_{L\mu \lambda }^0(k)\Pi _L^{\lambda
\rho }(k)D_{L\rho \nu }(k).  \eqnum{4.18}
\end{equation}
Solving the equations (4.17) and (4.18), one can get 
\begin{equation}
iD_{\mu \nu }^{ab}(k)=-i\delta ^{ab}\{\frac{g_{\mu \nu }-k_\mu k_\nu /k^2}{%
k^2+\Pi _T(k^2)-M^2+i\varepsilon }+\frac{\alpha k_\mu k_\nu /k^2}{k^2+\alpha
\Pi _L(k^2)-\sigma ^2+i\varepsilon }\}.  \eqnum{4.19}
\end{equation}
With setting

\begin{equation}
\Pi _T(k^2)=k^2\Pi _1(k^2)+M^2\Pi _2(k^2)  \eqnum{4.20}
\end{equation}
which follows from the Lorentz-covariance of the operator $\Pi _T(k^2)$ and

\begin{equation}
\alpha \Pi _L(k^2)=k^2\hat \Pi _L(k^2),  \eqnum{4.21}
\end{equation}
Eq. (4.19) will be written as 
\begin{equation}
iD_{\mu \nu }^{ab}(k)=-i\delta ^{ab}\{\frac{g_{\mu \nu }-k_\mu k_\nu /k^2}{%
k^2[1+\Pi _1(k^2)]-M^2[1-\Pi _2(k^2)]+i\varepsilon }+\frac{\alpha k_\mu
k_\nu /k^2}{k^2[1+\hat \Pi _L(k^2)]-\sigma ^2+i\varepsilon }.  \eqnum{4.22}
\end{equation}
We would like to note that the expressions given in Eqs. (4.12), (4.20) and
(4.21) can be verified by practical calculations and are important for
renormalizations of the propagators and the gluon mass.

Substitution of Eqs. (4.11) and (4.22) into Eq. (4.8) yields 
\begin{equation}
\hat \Pi _L(k^2)=\frac{\sigma ^2\hat \Omega (k^2)}{k^2[1+\hat \Omega (k^2)]}.
\eqnum{4.23}
\end{equation}
From this relation, we see, either in the Landau gauge or in the zero-mass
limit, the $\hat \Pi _L(k^2)$ vanishes.

Now let us discuss the renormalization. The function $\hat \Omega (k^2)$ in
Eq. (4.11) and the functions $\Pi _1(k^2)$, $\Pi _2(k^2)$ and $\hat \Pi
_L(k^2)$ in Eq. (4.22) are generally divergent in higher order perturbative
calculations. According to the conventional procedure of renormalization,
the divergences included in the functions $\hat \Omega (k^2),$ $\Pi _1(k^2),$
$\Pi _2(k^2)$ and $\hat \Pi _L(k^2)$ may be subtracted at a renormalization
point, say, $k^2=\mu ^2$. Thus, we can write [13-17] 
\begin{equation}
\begin{array}{c}
\hat \Omega (k^2)=\hat \Omega (\mu ^2)+\hat \Omega ^c(k^2),\;\;\Pi
_1(k^2)=\Pi _1(\mu ^2)+\Pi _1^c(k^2), \\ 
\Pi _2(k^2)=\Pi _2(\mu ^2)+\Pi _2^c(k^2),\text{ }\hat \Pi _L(k^2)=\hat \Pi
_L(\mu ^2)+\hat \Pi _L^c(k^2)
\end{array}
\eqnum{4.24}
\end{equation}
where $\hat \Omega (\mu ^2)$, $\Pi _1(\mu ^2),$ $\Pi _2(\mu ^2),$ $\hat \Pi
_L(\mu ^2)$ and $\Omega ^c(k^2)$, $\Pi _1^c(k^2)$, $\Pi _2^c(k^2),$ $\hat \Pi
_L^c(k^2)$ are respectively the divergent parts and the finite parts of the
functions $\Omega (k^2)$, $\Pi _1(k^2)$, $\Pi _2(k^2)$ and $\hat \Pi _L(k^2)$%
. The divergent parts can be absorbed in the following renormalization
constants defined by [13-17] 
\begin{equation}
\begin{array}{c}
\tilde Z_3^{-1}=1+\hat \Omega (\mu ^2),\;\;Z_3^{-1}=1+\Pi _1(\mu ^2),\text{ }%
Z_3^{\prime -1}=1+\hat \Pi _L(\mu ^2), \\ 
Z_M^{-1}=\sqrt{Z_3[1-\Pi _2(\mu ^2)]}=\sqrt{[1-\Pi _1(\mu ^2)][1-\Pi _2(\mu
^2)]}
\end{array}
\eqnum{4.25}
\end{equation}
where $Z_3$ and $\tilde Z_3$ are the renormalization constants of gluon and
ghost particle propagators respectively, $Z_3^{\prime }$ is the additional
renormalization constant of the longitudinal part of gluon propagator and $%
Z_M$ is the renormalization constant of gluon mass. With the above
definitions of the renormalization constants, on inserting Eq. (4.24) into
Eqs. (4.11) and (4.22) , the ghost particle propagator and gluon propagator
can be renormalized, respectively, in such a manner 
\begin{equation}
i\Delta ^{ab}(k)=\tilde Z_3i\Delta _R^{ab}(k)  \eqnum{4.26}
\end{equation}
and 
\begin{equation}
iD_{\mu \nu }^{ab}(k)=Z_3iD_{R\mu \nu }^{~~ab}(k)  \eqnum{4.27}
\end{equation}
where 
\begin{equation}
i\Delta _R^{ab}(k)=\frac{-i\delta ^{ab}}{k^2[1+\Omega _R(k^2)]-\sigma
_R^2+i\varepsilon }  \eqnum{4.28}
\end{equation}
and 
\begin{equation}
iD_{R\mu \nu }^{ab}(k)=-i\delta ^{ab}\{\frac{g_{\mu \nu }-k_\mu k_\nu /k^2}{%
k^2-M_R^2+\Pi _R^T(k^2)+i\varepsilon }+\frac{Z_3^{\prime }\alpha _Rk_\mu
k_\nu /k^2}{k^2[1+\Pi _R^L(k^2)]-\overline{\sigma }_R^2+i\varepsilon }\} 
\eqnum{4.29}
\end{equation}
are the renormalized propagators in which $M_R,$ $\overline{\sigma }_R$ and $%
\widetilde{\sigma }_R$ are the renormalized masses, $\alpha _R$ is the
renormalized gauge parameter, $\Omega _R(k^2),\Pi _R^T(k^2)$ and $\Pi
_R^L(k^2)$ denote the finite corrections coming from the loop diagrams. They
are defined as 
\begin{equation}
\begin{array}{c}
M_R=Z_M^{-1}M,\;\alpha _R=Z_3^{-1}\alpha ,\;\overline{\sigma }_R=\sqrt{%
Z_3^{^{\prime }}}\sigma ,\text{ }\sigma _R=\sqrt{\widetilde{Z}_3}\sigma , \\ 
\Omega _R(k^2)=\tilde Z_3\hat \Omega ^c(k^2),\text{ }\Pi
_R^T(k^2)=Z_3[k^2\Pi _1^c(k^2)+M^2\Pi _2^c(k^2)],\;\Pi _R^L(k^2)=Z_3^{\prime
}\hat \Pi _L^c(k^2).
\end{array}
\eqnum{4.30}
\end{equation}
The finite corrections above are zero at the renormalization point $\mu $.
As we see from Eq. (4.29), the longitudinal part of the gluon propagator,
except for in the Landau gauge, needs to be renormalized and has an extra
renormalization constant ${Z}_3^{\prime }$. This fact coincides with the
general property of the massive vector boson propagator (see Ref. (16),
Chap.V). From Eqs. (4.23)-(4.25) , it is easy to find that the longitudinal
part in Eq. ( 4.22) can be renormalized as 
\begin{equation}
\frac \alpha {k^2[1+\hat \Pi _L(k^2)]-\sigma ^2+i\varepsilon }=Z_3\alpha
_R[1+\Omega _R(k^2)]\Delta _R(k^2)  \eqnum{4.31}
\end{equation}
where 
\begin{equation}
\Delta _R(k^2)=\frac 1{k^2[1+\Omega _R(k^2)]-\sigma _R^2+i\varepsilon } 
\eqnum{4.32}
\end{equation}
which appears in Eq. (4.28) and the renormalization constant $Z_3^{\prime }$
can be expressed as 
\begin{equation}
Z_3^{\prime }=[1+\frac{\sigma _R^2}{\mu ^2}\frac{(1-\tilde Z_3)}{\tilde Z_3}%
]^{-1}.  \eqnum{4.33}
\end{equation}
If choosing $\mu =\sigma _R$, we have

\begin{equation}
Z_3^{\prime }=\tilde Z_3.  \eqnum{4.34}
\end{equation}

\section{Gluon three-line vertex}

The aim of this section is to derive the W-T identity satisfied by the gluon
three-line proper vertex and discuss its renormalization. For this purpose,
we first derive a W-T identity satisfied by the gluon three-point Green
function. Let us begin with the derivation from the W-T identity in Eq.
(4.1) and the ghost equation in Eq. (4.2). By taking successive
differentiations of Eq. (4.1) with respect to the sources $J_\nu ^b(y)$ and $%
J_\lambda ^c(z)$ and then setting the sources to vanish, one may obtain the
W-T identity obeyed by the gluon three-point Green function which is written
in the operator form as follows 
\begin{eqnarray}
\frac 1\alpha \partial _x^\mu G_{\mu \nu \lambda }^{abc}(x,y,z)
&=&<0^{+}|T^{*}[\hat {\bar C^a}(x)\hat D_\nu ^{bd}(y)\hat C^d(y)\hat A%
_\lambda ^c(z)]|0^{-}>  \nonumber \\
+ &<&0^{+}|T^{*}[\hat {\bar C^a}(x)\hat A_\nu ^b(y)\hat D_\lambda ^{cd}(z)%
\hat C^d(z)]|0^{-}>  \eqnum{5.1}
\end{eqnarray}
where 
\begin{equation}
G_{\mu \nu \lambda }^{abc}(x,y,z)=<0^{+}|T[\hat A_\mu ^a(x)\hat A_\nu ^b(y)%
\hat A_\lambda ^c(z)]|0^{-}>  \eqnum{5.2}
\end{equation}
is the three-point Green function mentioned above. The identity in Eq. (5.1)
will be simplified by a ghost equation which may be derived by
differentiating Eq. (4.2) with respect to the source $J_\lambda ^c(z)$ 
\begin{equation}
\begin{array}{c}
\partial _x^\mu <0^{+}|T^{*}\{\hat D_\mu ^{ad}(x)\hat C^d(x)\hat {\bar C^b}%
(y)\hat A_\lambda ^c(z)\}|0^{-}> \\ 
+{\sigma }^2<0^{+}|T[\hat C^a(x)\hat {\bar C^b}(y)\hat A_\lambda
^c(z)]|0^{-}]>=0.
\end{array}
\eqnum{5.3}
\end{equation}
Taking derivatives of Eq. (5.1) with respect to $y$ and $z$ and employing
Eq. (5.3), we get 
\begin{equation}
\partial _x^\mu \partial _y^\nu \partial _z^\lambda G_{\mu \nu \lambda
}^{abc}(x,y,z)=\alpha \sigma ^2\{\partial _y^\nu G_{~~\nu
}^{cab}(z,x,y)+\partial _z^\lambda G_{~~\lambda }^{bac}(y,x,z)\}  \eqnum{5.4}
\end{equation}
where 
\begin{equation}
G_{~~\mu }^{abc}(x,y,z)=<0^{+}|T\{\hat C^a(x)\hat {\bar C^b}(y)\hat A_\mu
^c(z)\}|0^{-}>.  \eqnum{5.5}
\end{equation}
In the Landau gauge or the zero-mass limit ($\sigma =0$), Eq. (5.4) reduces
to 
\begin{equation}
\partial _x^\mu \partial _y^\nu \partial _z^\lambda G_{\mu \nu \lambda
}^{abc}(x,y,z)=0  \eqnum{5.6}
\end{equation}
which shows the transversity of the Green function. From Eq. (5.4), we may
derive a W-T identity for the gluon three-line vertex. For this purpose, it
is necessary to use the following one-particle-irreducible decompositions of
the Green functions which can easily be obtained by the well-known procedure
[13-16] 
\begin{equation}
\begin{array}{c}
G_{\mu \nu \lambda }^{abc}(x,y,z)=\int d^4x^{\prime }d^4y^{\prime
}d^4z^{\prime }iD_{\mu \mu ^{\prime }}^{aa^{\prime }}(x-x^{\prime }) \\ 
\times iD_{\nu \nu ^{\prime }}^{bb^{\prime }}(y-y^{\prime })iD_{\lambda
\lambda ^{\prime }}^{cc^{\prime }}(z-z^{\prime })\Gamma _{a^{\prime
}b^{\prime }c^{\prime }}^{\mu ^{\prime }\nu ^{\prime }\lambda ^{\prime
}}(x^{\prime },y^{\prime },z^{\prime })
\end{array}
\eqnum{5.7}
\end{equation}
and 
\begin{equation}
\begin{array}{c}
G_{~~\,\nu }^{abc}(x,y,z)=\int d^4x^{\prime }d^4y^{\prime }d^4z^{\prime
}i\Delta ^{aa^{\prime }}(x-x^{\prime })\Gamma ^{a^{\prime }b^{\prime
}c^{\prime },\nu ^{\prime }}(x^{\prime },y^{\prime },z^{\prime }) \\ 
\times i\Delta ^{b^{\prime }b}(y^{\prime }-y)iD_{\nu ^{\prime }\nu
}^{c^{\prime }c}(z^{\prime }-z)
\end{array}
\eqnum{5.8}
\end{equation}
where $iD_{\mu \mu ^{\prime }}^{aa^{\prime }}(x-x^{\prime })$ and $i\Delta
^{aa^{\prime }}(x-x^{\prime })$ are respectively the gluon and the ghost
particle propagators discussed in the preceding section, $\Gamma _{abc}^{\mu
\nu \lambda }(x,y,z)$ and $\Gamma _{~~\lambda }^{abc}(x,y,z)$ are the
three-line gluon proper vertex and the three-line ghost-gluon proper vertex
respectively. They are defined as [13-16] 
\begin{equation}
\Gamma _{abc}^{\mu \nu \lambda }(x,y,z)=i\frac{\delta ^3\Gamma }{\delta
A_\mu ^a(x)\delta A_\nu ^b(y)\delta A_\lambda ^c(z)}|_{J=0}  \eqnum{5.9}
\end{equation}
and 
\begin{equation}
\Gamma _{~~\,\lambda }^{abc}(x,y,z)=\frac{\delta ^3\Gamma }{i\delta \bar C%
^a(x)\delta C^b(y)\delta A^{c\lambda }(z)}|_{J=0}  \eqnum{5.10}
\end{equation}
where $J$ stands for all the external sources. Substituting Eqs. (5.7) and
(5.8) into Eq. (5.4) and transforming Eq. (5.4) into the momentum space, one
can derive an identity which establishes the relation between the
longitudinal part of three-line gluon vertex and the three-line ghost-gluon
vertex as follows 
\begin{equation}
\begin{array}{c}
p^\mu q^\nu k^\lambda \Lambda _{\mu \nu \lambda }^{abc}(p,q,k)=-\frac{\sigma
^2}\alpha \chi (p^2)[\chi (k^2)q^\nu \Lambda _{~~\nu }^{cab}(k,p,q) \\ 
+\chi (q^2)k^\lambda \Lambda _{~~\,\lambda }^{bac}(q,p,k)]
\end{array}
\eqnum{5.11}
\end{equation}
where we have defined 
\begin{equation}
\begin{array}{c}
\Gamma _{\mu \nu \lambda }^{abc}(p,q,k)=(2\pi )^4\delta ^4(p+q+k)\Lambda
_{\mu \nu \lambda }^{abc}(p,q,k), \\ 
\Gamma _{~~\,\lambda }^{abc}(p,q,k)=(2\pi )^4\delta ^4(p+q+k)\Lambda
_{~~\lambda }^{abc}(p,q,k)
\end{array}
\eqnum{5.12}
\end{equation}
and 
\begin{equation}
\begin{array}{c}
\chi (p^2)=\{k^2[1+\hat \Pi _L(p^2)]-\sigma ^2+i\varepsilon \}\{k^2[1+\hat 
\Omega (p^2)]-\sigma ^2+i\varepsilon \}^{-1} \\ 
=[1+\widehat{\Omega }(k^2)]^{-1}
\end{array}
\eqnum{5.13}
\end{equation}
here $\hat \Pi _L(p^2)$ and $\hat \Omega (p^2)$ are the self-energies
defined in Eqs. (4.12) and (4.21). The second equality is obtained by
inserting the relation in Eq. (4.23) into the first equality.

Obviously, in the Landau gauge, Eq. (5.11) reduces to 
\begin{equation}
p^\mu q^\nu k^\lambda \Lambda _{\mu \nu \lambda }^{abc}(p,q,k)=0 
\eqnum{5.14}
\end{equation}
which implies that the vertex is transverse in this case. In the lowest
order approximation, owing to 
\begin{equation}
\chi (p^2)=1  \eqnum{5.15}
\end{equation}
and 
\begin{equation}
\Lambda _{~~~~~\;\mu }^{(0)abc}(p,q,k)=gf^{abc}p_\mu  \eqnum{5.16}
\end{equation}
where $f^{abc}$ are the structure constants of the gauge group, the right
hand side (RHS) of Eq. (5.11) vanishes, therefore, we have 
\begin{equation}
p^\mu q^\nu k^\lambda \Lambda _{~~~\mu \nu \lambda }^{(0)abc}(p,q,k)=0. 
\eqnum{5.17}
\end{equation}
This result is consistent with that for the bare three-line gluon vertex
given by the Feynman rule.

Now, let us discuss renormalization of the three-line gluon vertex. From the
renormalization of the gluon and ghost particle propagators described in
Eqs. (4.26) and (4.27) and the definitions of the propagators written in
Eqs. (4.6) and (4.7), one can see 
\begin{equation}
\begin{array}{c}
A^{a\mu }(x)=\sqrt{Z_3}A_R^{a\mu }(x), \\ 
C^a(x)=\sqrt{\tilde Z_3}C_R^a(x),\text{ }\bar C^a(x)=\sqrt{\tilde Z_3}\bar C%
_R^a(x)
\end{array}
\eqnum{5.18}
\end{equation}
(hereafter the subscript $R$ marks renormalized quantities). According to
above relations and the definitions given in Eqs. (5.9), (5.10) and (5.12),
we find 
\begin{equation}
\begin{array}{c}
\Lambda _{\mu \nu \lambda }^{abc}(p,q,k)=Z_3^{-3/2}\Lambda _{R\mu \nu
\lambda }^{~~abc}(p,q,k), \\ 
\Lambda _{~~\ \lambda }^{abc}(p,q,k)=\tilde Z_3^{-1}Z_3^{-1/2}\Lambda
_{R~~\lambda }^{~~abc}(p,q,k).
\end{array}
\eqnum{5.19}
\end{equation}
Applying these relations, the renormalized version of the identity written
in Eq. (5.11) will be 
\begin{equation}
\begin{array}{c}
p^\mu q^\nu k^\lambda \Lambda _{R\mu \nu \lambda }^{~~abc}(p,q,k)=-\frac{%
\sigma _R^2}{\alpha _R}{\chi }_R(p^2)[{\chi _R}(k^2)q^\nu \Lambda _{R~~\nu
}^{~~cab}(k,p,q) \\ 
+\chi _R(q^2)k^\lambda \Lambda _{R~~\lambda }^{~~bac}(q,p,k)]
\end{array}
\eqnum{5.20}
\end{equation}
where $\alpha _R$ and $\widetilde{\sigma }_R$ were defined in Eq. (4.30) and 
\begin{equation}
\chi _R(k^2)=\frac 1{1+\Omega _R(k^2)]}  \eqnum{5.21}
\end{equation}
is the renormalized expression of the function $\chi (k^2)$. In the above,
we have considered 
\begin{equation}
\chi (k^2)=\widetilde{Z}_3\chi _R(k^2)  \eqnum{5.22}
\end{equation}
which follows from $\hat \Omega (k^2)=\hat \Omega (\mu ^2)+\hat \Omega
^c(k^2),$ $\tilde Z_3^{-1}=1+\hat \Omega (\mu ^2)$ and$\;\Omega _R(k^2)=%
\tilde Z_3\hat \Omega ^c(k^2)$ defined in the preceding section. At the
renormalization point chosen to be $p^2=q^2=k^2=\mu ^2$, we see, $\chi
_R(\mu ^2)=1$. In this case, the renormalized ghost-gluon vertex takes the
form of the bare vertex so that the RHS of Eq. (5.20) vanishes, therefore,
we have 
\begin{equation}
p^\mu q^\nu k^\lambda \Lambda _{R\mu \nu \lambda
}^{~~abc}(p,q,k)|_{p^2=q^2=k^2=\mu ^2}=0.  \eqnum{5.23}
\end{equation}

Ordinarily, one is interested in discussing the renormalization of such
three-line vertices that they are defined from the vertices defined in Eqs.
(5.9) and (5.10) by extracting a coupling constant $g$. These vertices are
denoted by $\tilde \Lambda _{\mu \nu \lambda }^{abc}(p,q,k)$ and $\tilde 
\Lambda _{~~\lambda }^{abc}(p,q,k)$. Commonly, they are renormalized in such
a fashion [13-17] 
\begin{equation}
\begin{array}{c}
\tilde \Lambda _{\mu \nu \lambda }^{abc}(p,q,k)=Z_1^{-1}\widetilde{\Lambda }%
_{R\mu \nu \lambda }^{abc}(p,q,k), \\ 
\tilde \Lambda _{~~\lambda }^{abc}(p,q,k)=\widetilde{Z}_1^{-1}\widetilde{%
\Lambda }_{R~~\lambda }^{abc}(p,q,k)
\end{array}
\eqnum{5.24}
\end{equation}
where $Z_1$ and $\tilde Z_1$ are referred to as the renormalization
constants for the gluon three-line vertex and the ghost-gluon vertex,
respectively. It is clear that the W-T identity shown in Eq. (5.11) also
holds for the vertices $\tilde \Lambda _{\mu \nu \lambda }^{abc}(p,q,k)$ and 
$\tilde \Lambda _{~~\lambda }^{abc}(p,q,k)$. So, when the vertices $\Lambda
_{\mu \nu \lambda }^{abc}(p,q,k)$ and $\Lambda _{~~\lambda }^{abc}(p,q,k)$
in Eqs. (5.11) are replaced by $\tilde \Lambda _{\mu \nu \lambda
}^{abc}(p,q,k)$ and $\tilde \Lambda _{~~\lambda }^{abc}(p,q,k)$ respectively
and then Eq. (5.24) is inserted to such an identity, we obtain a
renormalized version of the identity as follows 
\begin{equation}
\begin{array}{c}
p^\mu q^\nu k^\lambda \tilde \Lambda _{R\mu \nu \lambda }^{~~abc}(p,q,k)=-%
\frac{Z_1\tilde Z_3}{Z_3\tilde Z_1}\frac{\widetilde{\sigma }_R^2}{\alpha _R}%
\chi _R(p^2)[\chi _R(k^2) \\ 
\times q^\nu \tilde \Lambda _{R~~\nu }^{~~cab}(k,p,q)+\chi _R(q^2)k^\lambda 
\tilde \Lambda _{R~~\lambda }^{~~bac}(q,p,k)].
\end{array}
\eqnum{5.25}
\end{equation}
When multiplying the both sides of Eq. (5.25) with a renormalized coupling
constant $g_R$ and absorbing it in the vertices, noticing 
\begin{equation}
\begin{array}{c}
\Lambda _{R\mu \nu \lambda }^{abc}(p,q,k)=g_R\widetilde{\Lambda }_{R\mu \nu
\lambda }^{abc}(p,q,k), \\ 
\Lambda _{R~~\lambda }^{abc}(p,q,k)=g_R\widetilde{\Lambda }_{R~~\lambda
}^{abc}(p,q,k),
\end{array}
\eqnum{5.26}
\end{equation}
we have 
\begin{equation}
\begin{array}{c}
p^\mu q^\nu k^\lambda \Lambda _{R\mu \nu \lambda }^{~~abc}(p,q,k)=-\frac{Z_1%
\tilde Z_3}{Z_3\tilde Z_1}\frac{\sigma _R^2}{\alpha _R}\chi _R(p^2)[\chi
_R(k^2) \\ 
\times q^\nu \Lambda _{R~~\nu }^{~~cab}(k,p,q)+\chi _R(q^2)k^\lambda \Lambda
_{R~~\lambda }^{~~bac}(q,p,k)].
\end{array}
\eqnum{5.27}
\end{equation}
In comparison of Eq. (5.27) with Eq. (5.20), we see, except for the factor $%
Z_1\tilde Z_3Z_3^{-1}\tilde Z_1^{-1}$, the both identities are identical to
each other. From this observation, we deduce 
\begin{equation}
\frac{Z_1}{Z_3}=\frac{\tilde Z_1}{\tilde Z_3}.  \eqnum{5.28}
\end{equation}
This is the S-T identity which coincides with the one given in the massless
QCD [19, 20].

\section{Gluon four-line vertex}

By the similar procedure as deriving Eqs. (5.1) and (5.3), the W-T identity
obeyed by the gluon four-point Green function may be derived by
differentiating Eq. (4.1) with respect to the sources $J_\mu ^b(y),J_\lambda
^c(z)$ and $J_\tau ^d(u)$. The result represented in the operator form is as
follows 
\begin{equation}
\begin{array}{c}
\frac 1\alpha \partial _x^\mu G_{\mu \nu \lambda \tau }^{abcd}(x,y,z,u) \\ 
=<0^{+}|T^{*}[\hat {\bar C^a}(x)\hat D_\nu ^{be}(y)\hat C^e(y)\hat A_\lambda
^c(z)\hat A_\tau ^d(u)]|0^{-}> \\ 
+<0^{+}|T^{*}[\hat {\bar C^a}(x)\hat A_\nu ^b(y)\hat D_\lambda ^{ce}(z)\hat C%
^e(z)\hat A_\tau ^d(u)]|0^{-}> \\ 
+<0^{+}|T^{*}[\hat {\bar C^a}(x)\hat A_\nu ^b(y)\hat A_\lambda ^c(z)\hat D%
_\tau ^{de}(u)\hat C^e(u)]|0^{-}>
\end{array}
\eqnum{6.1}
\end{equation}
where 
\begin{equation}
G_{\mu \nu \lambda \tau }^{abcd}(x,y,z,u)=<0^{+}|T[\hat A_\mu ^a(x)\hat A%
_\nu ^b(y)\hat A_\lambda ^c(z)\hat A_\tau ^d(u)]|0^{-}>  \eqnum{6.2}
\end{equation}
is the gluon four-point Green function. The accompanying ghost equation may
be obtained by differentiating Eq. (4.2) with respect to the sources $%
J_\lambda ^c(z)$ and $J_\tau ^d(u)$. The result is 
\begin{equation}
\begin{array}{c}
\partial _x^\mu <0^{+}|T^{*}[\hat D_\mu ^{ae}(x)\hat C^e(x)\hat {\bar C^b}(y)%
\hat A_\lambda ^c(z)\hat A_\tau ^d(u)]|0^{-}> \\ 
+{\sigma }^2G_{~~\lambda \tau }^{abcd}(x,y,z,u)=-\delta ^{ab}\delta
^4(x-y)D_{\lambda \tau }^{cd}(z-u)
\end{array}
\eqnum{6.3}
\end{equation}
where 
\begin{equation}
G_{~~\lambda \tau }^{abcd}(x,y,z,u)=<0^{+}|T[\hat C^a(x)\hat {\bar C^b}(y)%
\hat A_\lambda ^c(z)\hat A_\tau ^d(u)]|0^{-}>  \eqnum{6.4}
\end{equation}
is the four-point gluon-ghost particle Green function. Differentiation of
Eq. (6.1) with respect to the coordinates $y$, $z$ and $u$ and use of Eq.
(6.3) lead to 
\begin{equation}
\begin{array}{c}
\frac 1\alpha \partial _x^\mu \partial _y^\nu \partial _z^\lambda \partial
_u^\tau G_{\mu \nu \lambda \tau }^{abcd}(x,y,z,u)=\delta ^{ab}\delta
^4(x-y)\partial _z^\lambda \partial _u^\tau D_{\lambda \tau }^{cd}(z-u) \\ 
+\delta ^{ac}\delta ^4(x-z)\partial _y^\nu \partial _u^\tau D_{\nu \tau
}^{bd}(y-u)+\delta ^{ad}\delta ^4(x-u)\partial _y^\nu \partial _z^\lambda
D_{\nu \lambda }^{bc}(y-z) \\ 
+\sigma ^2\{\partial _z^\lambda \partial _u^\tau G_{~~\lambda \tau
}^{bacd}(y,x,z,u)+\partial _y^\nu \partial _u^\tau G_{~~\nu \tau
}^{cabd}(z,x,y,u) \\ 
+\partial _y^\nu \partial _z^\lambda G_{~~\nu \lambda }^{dabc}(u,x,y,z)\}.
\end{array}
\eqnum{6.5}
\end{equation}

It is noted that the four-point Green functions appearing in the above
equations are unconnected. Their decompositions to connected Green functions
are not difficult to be found by making use of the relation between the
generating functionals $Z$ for the full Green functions and $W$ for the
connected Green functions as written in Eq. (3.9). The result is 
\begin{equation}
\begin{array}{c}
G_{\mu \nu \lambda \tau }^{abcd}(x,y,z,u)=G_{\mu \nu \lambda \tau
}^{abcd}(x,y,z,u)_c-D_{\mu \nu }^{ab}(x-y)D_{\lambda \tau }^{cd}(z-u) \\ 
-D_{\mu \lambda }^{ac}(x-z)D_{\nu \tau }^{bd}(y-u)-D_{\mu \tau
}^{ad}(x-u)D_{\nu \lambda }^{bc}(y-z)
\end{array}
\eqnum{6.6}
\end{equation}
and 
\begin{equation}
G_{~~\lambda \tau }^{abcd}(x,y,z,u)=G_{~~\,\lambda \tau
}^{abcd}(x,y,z,u)_c-\Delta ^{ab}(x-y)D_{\lambda \tau }^{cd}(z-u). 
\eqnum{6.7}
\end{equation}
The first terms marked by the subscript ''$c$'' in Eqs. ( 6.6) and (6.7) are
connected Green functions. When inserting Eqs. (6.6) and (6.7) into Eq.
(6.5) and using the W-T identity in Eq. (4.5), one may find 
\begin{equation}
\begin{array}{c}
\partial _x^\mu \partial _y^\nu \partial _z^\lambda \partial _u^\tau G_{\mu
\nu \lambda \tau }^{abcd}(x,y,z,u)_c=\alpha \sigma ^2\{\partial _y^\nu
\partial _z^\lambda G_{~~\nu \lambda }^{dabc}(u,x,y,z)_c \\ 
+\partial _y^\nu \partial _u^\tau G_{~~\nu \tau }^{cabd}(z,x,y,u)_c+\partial
_z^\lambda \partial _u^\tau G_{~~\lambda \tau }^{bacd}(y,x,z,u)_c\}.
\end{array}
\eqnum{6.8}
\end{equation}
This is the W-T identity satisfied by the connected four-point Green
functions. In the Landau gauge, we have 
\begin{equation}
\partial _x^\mu \partial _y^\nu \partial _z^\lambda \partial _u^\tau G_{\mu
\nu \lambda \tau }^{abcd}(x,y,z,u)_c=0  \eqnum{6.9}
\end{equation}
which shows the transversity of the Green function.

The W-T identity for the four-line proper gluon vertex may be derived from
Eq. (6.8) with the help of the following one-particle-irreducible
decompositions of the connected Green functions which can easily be found by
the standard procedure [13-16]. 
\begin{equation}
\begin{array}{c}
~G_{\mu \nu \lambda \tau }^{abcd}(x_1,x_2,x_3,x_4)_c \\ 
=\int \prod\limits_{i=1}^4d^4y_iD_{\mu \mu ^{\prime }}^{aa^{\prime
}}(x_1-y_1)D_{\nu \nu ^{\prime }}^{bb^{\prime }}(x_2-y_2)\Gamma _{a^{\prime
}b^{\prime }c^{\prime }d^{\prime }}^{\mu ^{\prime }\nu ^{\prime }\lambda
^{\prime }\tau ^{\prime }}(y_1,y_2,y_3,y_4) \\ 
\times D_{\lambda ^{\prime }\lambda }^{c^{\prime }c}(y_3-x_3)D_{\tau
^{\prime }\tau }^{d^{\prime }d}(y_4-x_4) \\ 
+i\int \prod\limits_{i=1}^4d^4y_id^4z_i\{D_{\mu \mu ^{\prime }}^{aa^{\prime
}}(x_1-y_1)D_{\nu \nu ^{\prime }}^{bb^{\prime }}(x_2-y_2)\Gamma _{a^{\prime
}b^{\prime }e}^{\mu ^{\prime }\nu ^{\prime }\rho }(y_1,y_2,y_3) \\ 
\times D_{\rho \rho ^{\prime }}^{ee^{\prime }}(y_3-z_1)\Gamma _{e^{\prime
}c^{\prime }d^{\prime }}^{\rho ^{\prime }\lambda ^{\prime }\tau ^{\prime
}}(z_1,z_2,z_3)D_{\lambda ^{\prime }\lambda }^{c^{\prime }c}(z_2-x_3)D_{\tau
^{\prime }\tau }^{d^{\prime }d}(z_3-x_4) \\ 
+D_{\mu \mu ^{\prime }}^{aa^{\prime }}(x_1-y_1)D_{\lambda \lambda ^{\prime
}}^{cc^{\prime }}(x_3-y_2)\Gamma _{a^{\prime }c^{\prime }e}^{\mu ^{\prime
}\lambda ^{\prime }\rho }(y_1,y_2,y_3)D_{\rho \rho ^{\prime }}^{ee^{\prime
}}(y_3-z_1) \\ 
\times \Gamma _{e^{\prime }b^{\prime }d^{\prime }}^{\rho ^{\prime }\nu
^{\prime }\tau ^{\prime }}(z_1,z_2,z_3)D_{\nu ^{\prime }\nu }^{b^{\prime
}b}(z_2-x_2)D_{\tau ^{\prime }\tau }^{d^{\prime }d}(z_3-x_4) \\ 
+D_{\nu \nu ^{\prime }}^{bb^{\prime }}(x_2-y_1)D_{\lambda \lambda ^{\prime
}}^{cc^{\prime }}(x_3-y_2)\Gamma _{b^{\prime }c^{\prime }e}^{\nu ^{\prime
}\lambda ^{\prime }\rho }(y_1,y_2,y_3)D_{\rho \rho ^{\prime }}^{ee^{\prime
}}(y_3-z_1) \\ 
\times \Gamma _{e^{\prime }a^{\prime }d^{\prime }}^{\rho ^{\prime }\mu
^{\prime }\tau ^{\prime }}(z_1,z_2,z_3)D_{\mu ^{\prime }\mu }^{a^{\prime
}a}(z_2-x_1)D_{\tau ^{\prime }\tau }^{d^{\prime }d}(z_3-x_4)\}
\end{array}
\eqnum{6.10}
\end{equation}
and 
\begin{equation}
\begin{array}{c}
G_{~~\lambda \tau }^{abcd}(x_1,x_2,x_3,x_4)_c \\ 
=\int \prod\limits_{i=1}^4d^4y_i\Delta ^{aa^{\prime }}(x_1-y_1)\Gamma
_{a^{\prime }b^{\prime }c^{\prime }d^{\prime }}^{~~~~\lambda ^{\prime }\tau
^{\prime }}(y_1,y_2,y_3,y_4)\Delta ^{b^{\prime }b}(y_2-x_2) \\ 
\times D_{\lambda ^{\prime }\lambda }^{c^{\prime }c}(y_3-x_3)D_{\tau
^{\prime }\tau }^{d^{\prime }d}(y_4-x_4) \\ 
+i\int \prod\limits_{i=1}^4d^4y_id^4z_i\{\Delta ^{aa^{\prime
}}(x_1-y_1)\Gamma _{a^{\prime }ed^{\prime }}^{~~~\tau ^{\prime
}}(y_1,y_2,y_3)\Delta ^{ee^{\prime }}(y_2-z_1) \\ 
\times D_{\tau ^{\prime }\tau }^{d^{\prime }d}(y_3-x_4)\Gamma _{e^{\prime
}b^{\prime }c^{\prime }}^{~~~~\lambda ^{\prime }}(z_1,z_2,z_3)\Delta
^{b^{\prime }b}(z_2-x_2)D_{\lambda ^{\prime }\lambda }^{c^{\prime
}c}(z_3-x_3) \\ 
+\Delta ^{aa^{\prime }}(x_1-y_1)\Gamma _{a^{\prime }ec^{\prime
}}^{~~~\lambda ^{\prime }}(y_1,y_2,y_3)\Delta ^{ee^{\prime
}}(y_2-z_1)D_{\lambda ^{\prime }\lambda }^{c^{\prime }c}(y_3-x_3) \\ 
\times \Gamma _{e^{\prime }b^{\prime }d^{\prime }}^{~~~\tau ^{\prime
}}(z_1,z_2,z_3)\Delta ^{b^{\prime }b}(z_2-x_2)D_{\tau ^{\prime }\tau
}^{d^{\prime }d}(z_3-x_4) \\ 
+\Delta ^{aa^{\prime }}(x_1-y_1)\Gamma _{a^{\prime }b^{\prime }e}^{~~~\rho
}(y_1,y_2,y_3)\Delta ^{b^{\prime }b}(y_2-x_2)D_{\rho \rho ^{\prime
}}^{ee^{\prime }}(y_3-z_1) \\ 
\times \Gamma _{e^{\prime }c^{\prime }d^{\prime }}^{\rho ^{\prime }\lambda
^{\prime }\tau ^{\prime }}(z_1,z_2,z_3)D_{\lambda ^{\prime }\lambda
}^{c^{\prime }c}(z_2-x_3)D_{\tau ^{\prime }\tau }^{d^{\prime }d}(z_3-x_4)\}
\end{array}
\eqnum{6.11}
\end{equation}
where $\Gamma _{\mu \nu \lambda \tau }^{abcd}(x_1,x_2,x_3,x_4)$ is the
four-line gluon proper vertex and $\Gamma _{~~\lambda \tau
}^{abcd}(x_1,x_2,x_3,x_4)$ is the four-line ghost-gluon proper vertex. They
are defined as [13-16] 
\begin{equation}
\begin{array}{c}
\Gamma _{\mu \nu \lambda \tau }^{abcd}(x_1,x_2,x_3,x_4)=i\frac{\delta
^4\Gamma }{\delta A^{a\mu }(x_1)\delta A^{b\nu }(x_2)\delta A^{c\lambda
}(x_3)\delta A^{d\tau }(x_4)}|_{J=0}, \\ 
\Gamma _{~~\lambda \tau }^{abcd}(x_1,x_2,x_3,x_4)=\frac{\delta ^4\Gamma }{%
i\delta \bar C^a(x_1)\delta C^b(x_2)\delta A^{c\lambda }(x_3)\delta A^{d\tau
}(x_4)}|_{J=0}.
\end{array}
\eqnum{6.12}
\end{equation}
When substituting Eqs. (6.10) and (6.11) into Eq. (6.8) and transforming Eq.
(6.8) into the momentum space, one can find the following identity satisfied
by the four-line proper gluon vertex 
\begin{equation}
\begin{array}{c}
k_1^\mu k_2^\nu k_3^\lambda k_4^\tau \Lambda _{\mu \nu \lambda \tau
}^{abcd}(k_1,k_2,k_3,k_4)=\Psi \left( 
\begin{array}{cccc}
a & b & c & d \\ 
k_1 & k_2 & k_3 & k_4
\end{array}
\right) \\ 
+\Psi \left( 
\begin{array}{cccc}
a & c & d & b \\ 
k_1 & k_3 & k_4 & k_2
\end{array}
\right) +\Psi \left( 
\begin{array}{cccc}
a & d & b & c \\ 
k_1 & k_4 & k_2 & k_3
\end{array}
\right)
\end{array}
\eqnum{6.13}
\end{equation}
where 
\begin{equation}
\begin{array}{c}
\Psi \left( 
\begin{array}{cccc}
a & b & c & d \\ 
k_1 & k_2 & k_3 & k_4
\end{array}
\right) \\ 
=-ik_1^\mu k_2^\nu \Lambda _{\mu \nu \sigma
}^{abe}(k_1,k_2,-(k_1+k_2))D_{ef}^{\sigma \rho }(k_1+k_2)k_3^\lambda
k_4^\tau \Lambda _{\rho \lambda \tau }^{fcd}(-(k_3+k_4),k_3,k_4) \\ 
+\frac{i\sigma ^2}\alpha \chi (k_1^2)\chi (k_2^2)[ik_3^\lambda k_4^\tau
\Lambda _{~~\lambda \tau }^{bacd}(k_2,k_1,k_3,k_4) \\ 
-\Lambda _{~~\sigma }^{bae}(k_2,k_1,-(k_1+k_2))D_{ef}^{\sigma \rho
}(k_1+k_2)k_3^\lambda k_4^\tau \Lambda _{\rho \lambda \tau
}^{fcd}(-(k_3+k_4),k_3,k_4) \\ 
-k_4^\tau \Lambda _{~~\tau }^{bed}(k_2,-(k_2+k_4),k_4)\Delta
^{ef}(k_2+k_4)k_3^\lambda \Lambda _{~~\lambda }^{fac}(-(k_1+k_3),k_1,k_3) \\ 
-k_3^\lambda \Lambda _{~~\lambda }^{bec}(k_2,-(k_2+k_3),k_3)\Delta
^{ef}(k_2+k_3)k_4^\tau \Lambda _{~~~\tau }^{fad}(-(k_1+k_4),k_1,k_4)].
\end{array}
\eqnum{6.14}
\end{equation}
The second and third terms in Eq .(6.13) can be written out from Eq. (6.14)
through cyclic permutations. In the above, we have defined 
\begin{equation}
\begin{array}{c}
\Gamma _{\mu \nu \lambda \tau }^{abcd}(k_1,k_2,k_3,k_4)=(2\pi )^4\delta
^4(\sum_{i=1}^4k_i)\Lambda _{\mu \nu \lambda \tau }^{abcd}(k_1,k_2,k_3,k_4),
\\ 
\Gamma _{~~\lambda \tau }^{abcd}(k_1,k_2,k_3,k_4)=(2\pi )^4\delta
^4(\sum_{i=1}^4k_i)\Lambda _{~~\lambda \tau }^{abcd}(k_1,k_2,k_3,k_4).
\end{array}
\eqnum{6.15}
\end{equation}
In the lowest order approximation, we have checked that except for the first
term in Eq. (6.14) which was encountered in the massless theory, the
remaining mass-dependent terms are cancelled out with the corresponding
terms contained in the second and third terms in Eq. (6.13). Therefore, the
identity in Eq. (6.13) leads to a result in the lowest order approximation
which is consistent with the Feynman rule.

The renormalization of the four-line vertices is similar to that for the
three-line vertices. From the definitions given in Eqs. (6.12), (6.15) and
(5.18), it is clearly seen that the four-line vertices should be
renormalized in such a manner 
\begin{equation}
\begin{array}{c}
\Lambda _{\mu \nu \lambda \tau }^{abcd}(k_1,k_2,k_3,k_4)=Z_3^{-2}\Lambda
_{R\mu \nu \lambda \tau }^{~~abcd}(k_1,k_2,k_3,k_4), \\ 
\Lambda _{~~\lambda \tau }^{abcd}(k_1,k_2,k_3,k_4)=\tilde Z%
_3^{-1}Z_3^{-1}\Lambda _{R\text{ }~~\lambda \tau }^{~~abcd}(k_1,k_2,k_3,k_4).
\end{array}
\eqnum{6.16}
\end{equation}
On inserting these relations into Eqs. (6.13) and (6.14), one can obtain a
renormalized identity similar to Eq. (5.20), that is 
\begin{equation}
\begin{array}{c}
k_1^\mu k_2^\nu k_3^\lambda k_4^\tau \Lambda _{R\mu \nu \lambda \tau }^{%
\text{ }abcd}(k_1,k_2,k_3,k_4)=\Psi _R\left( 
\begin{array}{cccc}
a & b & c & d \\ 
k_1 & k_2 & k_3 & k_4
\end{array}
\right) \\ 
+\Psi _R\left( 
\begin{array}{cccc}
a & c & d & b \\ 
k_1 & k_3 & k_4 & k_2
\end{array}
\right) +\Psi _R\left( 
\begin{array}{cccc}
a & d & b & c \\ 
k_1 & k_4 & k_2 & k_3
\end{array}
\right)
\end{array}
\eqnum{6.17}
\end{equation}
where 
\begin{equation}
\begin{array}{c}
\Psi _R\left( 
\begin{array}{cccc}
a & b & c & d \\ 
k_1 & k_2 & k_3 & k_4
\end{array}
\right) \\ 
=-ik_1^\mu k_2^\nu \Lambda _{R\mu \nu \sigma }^{\text{ }%
abe}(k_1,k_2,-(k_1+k_2))D_{Ref}^{\text{ }\sigma \rho }(k_1+k_2)k_3^\lambda
k_4^\tau \Lambda _{R\rho \lambda \tau }^{fcd}(-(k_3+k_4),k_3,k_4) \\ 
+\frac{i\sigma _R^2}{\alpha _R}\chi _R(k_1^2)\chi _R(k_2^2)[ik_3^\lambda
k_4^\tau \Lambda _{R~~\lambda \tau }^{bacd}(k_2,k_1,k_3,k_4) \\ 
-\Lambda _{R~\sigma }^{bae}(k_2,k_1,-(k_1+k_2))D_{Ref}^{\text{ }\sigma \rho
}(k_1+k_2)k_3^\lambda k_4^\tau \Lambda _{R\rho \lambda \tau }^{\text{ }%
fcd}(-(k_3+k_4),k_3,k_4) \\ 
-k_4^\tau \Lambda _{R~~\tau }^{bed}(k_2,-(k_2+k_4),k_4)\Delta
_R^{ef}(k_2+k_4)k_3^\lambda \Lambda _{R~~\lambda }^{\text{ }%
fac}(-(k_1+k_3),k_1,k_3) \\ 
-k_3^\lambda \Lambda _{R~\lambda }^{bec}(k_2,-(k_2+k_3),k_3)\Delta
_R^{ef}(k_2+k_3)k_4^\tau \Lambda _{R~\tau }^{fad}(-(k_1+k_4),k_1,k_4)].
\end{array}
\eqnum{6.18}
\end{equation}
We can also define vertices $\tilde \Lambda _{\mu \nu \lambda \tau
}^{abcd}(k_1,k_2,k_3,k_4)$ and $\tilde \Lambda _{~~\text{ }\lambda \tau
}^{abcd}(k_1,k_2,k_3,k_4)$ from the vertices $\Lambda _{\mu \nu \lambda \tau
}^{abcd}(k_1,k_2,k_3,k_4)$ and $\Lambda _{~~\lambda \tau
}^{abcd}(k_1,k_2,k_3,k_4)$ by taking out the coupling constant squared,
respectively. The renormalization of these vertices are usually defined by
[13-17] 
\begin{equation}
\begin{array}{c}
\tilde \Lambda _{\mu \nu \lambda \tau }^{abcd}(k_1,k_2,k_3,k_4)=Z_4^{-1}%
\tilde \Lambda _{R\mu \nu \lambda \tau }^{~~abcd}(k_1,k_2,k_3,k_4), \\ 
\tilde \Lambda _{~~\lambda \tau }^{abcd}(k_1,k_2,k_3,k_4)=\tilde Z_4^{-1}%
\tilde \Lambda _{R~~\lambda \tau }^{~~abcd}(k_1,k_2,k_3,k_4).
\end{array}
\eqnum{6.19}
\end{equation}
where $Z_4$ and $\tilde Z_4$ are the renormalization constants of the
four-line gluon and ghost-gluon vertices respectively. Obviously, the
identity in Eqs. (6.13) and (6.14) remains formally unchanged if we replace
all the vertices $\Lambda _i$ in the identity with the ones $\widetilde{%
\Lambda }_i$ . Substituting Eqs. (5.24), (6.19), (4.26) and (4.27) into such
an identity, one may write a renormalized identity similar to Eq. (5.25),
that is 
\begin{equation}
\begin{array}{c}
k_1^\mu k_2^\nu k_3^\lambda k_4^\tau \widetilde{\Lambda }_{R\mu \nu \lambda
\tau }^{\text{ }abcd}(k_1,k_2,k_3,k_4)=\widetilde{\Psi }_R\left( 
\begin{array}{cccc}
a & b & c & d \\ 
k_1 & k_2 & k_3 & k_4
\end{array}
\right) \\ 
+\widetilde{\Psi }_R\left( 
\begin{array}{cccc}
a & c & d & b \\ 
k_1 & k_3 & k_4 & k_2
\end{array}
\right) +\widetilde{\Psi }_R\left( 
\begin{array}{cccc}
a & d & b & c \\ 
k_1 & k_4 & k_2 & k_3
\end{array}
\right)
\end{array}
\eqnum{6.20}
\end{equation}
where 
\begin{equation}
\begin{array}{c}
\widetilde{\Psi }_R\left( 
\begin{array}{cccc}
a & b & c & d \\ 
k_1 & k_2 & k_3 & k_4
\end{array}
\right) \\ 
=\frac{Z_4Z_3}{Z_1^2}\{-ik_1^\mu k_2^\nu \widetilde{\Lambda }_{R\mu \nu
\sigma }^{\text{ }abe}(k_1,k_2,-(k_1+k_2))D_{Ref}^{\text{ }\sigma \rho
}(k_1+k_2)k_3^\lambda k_4^\tau \widetilde{\Lambda }_{R\rho \lambda \tau }^{%
\text{ }fcd}(-(k_3+k_4),k_3,k_4)\} \\ 
+\frac{i\sigma _R^2}{\alpha _R}\chi _R(k_1^2)\chi _R(k_2^2)\{\frac{%
\widetilde{Z}_3Z_4}{Z_3\tilde Z_4}ik_3^\lambda k_4^\tau \widetilde{\Lambda }%
_{R~~\lambda \tau }^{\text{ }bacd}(k_2,k_1,k_3,k_4) \\ 
-\frac{Z_4\widetilde{Z}_3}{Z_1\tilde Z_1}\widetilde{\Lambda }_{R~~\sigma }^{%
\text{ }bae}(k_2,k_1,-(k_1+k_2))D_{Ref}^{\text{ }\sigma \rho
}(k_1+k_2)k_3^\lambda k_4^\tau \widetilde{\Lambda }_{R\rho \lambda \tau }^{%
\text{ }fcd}(-(k_3+k_4),k_3,k_4) \\ 
-\frac{Z_4\widetilde{Z}_3^2}{Z_3\widetilde{Z}_1^2}[k_4^\tau \widetilde{%
\Lambda }_{R~\tau }^{bed}(k_2,-(k_2+k_4),k_4)\Delta
_R^{ef}(k_2+k_4)k_3^\lambda \widetilde{\Lambda }_{R~~\lambda }^{\text{ }%
fac}(-(k_1+k_3),k_1,k_3) \\ 
+k_3^\lambda \widetilde{\Lambda }_{R~\lambda
}^{bec}(k_2,-(k_2+k_3),k_3)\Delta _R^{ef}(k_2+k_3)k_4^\tau \widetilde{%
\Lambda }_{R~\tau }^{fad}(-(k_1+k_4),k_1,k_4)]\}.
\end{array}
\eqnum{6.21}
\end{equation}
Multiplying the both sides of Eqs. (6.20) and (6.21) by $g_R^2$, according
to the relations given in Eqs. (5.26) and in the following 
\begin{equation}
\begin{array}{c}
\Lambda _{R\mu \nu \lambda \tau }^{abcd}(k_{1,}k_2,k_3,k_4)=g_R^2\tilde 
\Lambda _{R\mu \nu \lambda \tau }^{~~abcd}(k_1,k_2,k_3,k_4), \\ 
\Lambda _{R~~\lambda \tau }^{abcd}(k_1,k_2,k_3,k_4)=g_R^2\widetilde{\Lambda }%
_{R~~\lambda \tau }^{~~abcd}(k_1,k_2,k_3,k_4),
\end{array}
\eqnum{6.22}
\end{equation}
we have an identity which is of the same form as the identity in Eqs. (6.20)
and (6.21) except that the vertices $\widetilde{\Lambda }_R^i$ in Eqs.
(6.20) and (6.21) are all replaced by the vertices $\Lambda _R^i$. Comparing
this identity with that written in Eqs. (6.17) and (6.18), one may find 
\begin{equation}
\frac{Z_3Z_4}{Z_1^2}=1,\frac{\widetilde{Z}_3Z_4}{Z_3\widetilde{Z}_4}=1,\frac{%
Z_4\widetilde{Z}_3}{Z_1\widetilde{Z}_1}=1,\frac{Z_4\widetilde{Z}_3^2}{Z_3%
\widetilde{Z}_1^2}=1  \eqnum{6.23}
\end{equation}
which lead to 
\begin{equation}
\frac{Z_1}{Z_3}=\frac{\widetilde{Z}_1}{\widetilde{Z}_3}=\frac{Z_4}{Z_1},%
\frac{Z_1}{\widetilde{Z}_1}=\frac{Z_3}{\widetilde{Z}_3}=\frac{Z_4}{%
\widetilde{Z}_4}  \eqnum{6.24}
\end{equation}
This just is the S-T identity which is consistent with that given in Refs.
(19) and (20) for the massless QCD.

\section{Quark-gluon vertex and quark propagator}

This section is used to derive the W-T identity for quark-gluon vertex and
discuss its renormalization. First we derive a W-T identity satisfied by the
quark-antiquark-gluon three-point Green function. This identity can easily
be derived by differentiating the W-T identity in Eq. (3.8) or (3.10) with
respect to the sources $\xi ^b(z)$, $\eta (y)$ and $\overline{\eta }(x)$ and
then setting all the sources to be zero. The result written in the operator
form is as follows 
\begin{equation}
\partial _z^\mu G_\mu ^a(x,y,z)=i\alpha
g[G_1^{ba}(x,y,y,z)T^b-T^bG_2^{ba}(x,y,x,z)]  \eqnum{7.1}
\end{equation}
where 
\begin{equation}
G_\mu ^a(x,y,z)=\left\langle 0^{+}\mid \widehat{\psi }(x)\widehat{\overline{%
\psi }}(y)\widehat{A}(z)\mid 0^{-}\right\rangle  \eqnum{7.2}
\end{equation}
is the quark-gluon three-point Green function, 
\begin{equation}
G_1^{ba}(x,y,z)=\left\langle 0^{+}\mid \widehat{\psi }(x)\widehat{\overline{%
\psi }}(y)\widehat{C}^b(y)\widehat{\overline{C}}^a(z)\mid 0^{-}\right\rangle
\eqnum{7.3}
\end{equation}
and 
\begin{equation}
G_2^{ba}(x,y,z)=\left\langle 0^{+}\mid \widehat{\psi }(x)\widehat{\overline{%
\psi }}(y)\widehat{C}^b(x)\widehat{\overline{C}}^a(z)\mid 0^{-}\right\rangle
\eqnum{7.4}
\end{equation}
are the quark-ghost particle mixed Green functions. The Green functions in
Eqs. (7.3) and (7.4) are connected because a quark field and a ghost field
are of a common coordinate.

The W-T identity for quark-gluon vertex can be derived from Eq. (7.1) with
the help of one-particle irreducible decompositions of the Green functions
shown in Eqs. (7.2)-(7.4). The decompositions can easily be obtained by the
standard procedure [13-16]. The results are given in the following. 
\begin{equation}
G_\mu ^a(x,y,z)=\int d^4x^{\prime }d^4y^{\prime }d^4z^{\prime
}iS_F(x-x^{\prime })\Gamma ^{b\nu }(x^{\prime },y^{\prime },z^{\prime
})iS_F(y^{\prime }-y)iD_{\nu \mu }^{ba}(z^{\prime }-z)  \eqnum{7.5}
\end{equation}
where $D_{\nu \mu }^{ba}(z^{\prime }-z)$ is the gluon propagator defined in
Eq. (4.6), 
\begin{equation}
iS_F(x-x^{\prime })=\left\langle 0^{+}\mid \widehat{\psi }(x)\widehat{%
\overline{\psi }}(x^{\prime })\mid 0^{-}\right\rangle  \eqnum{7.6}
\end{equation}
is the quark propagator and 
\begin{equation}
\Gamma ^{b\nu }(x^{\prime },y^{\prime },z^{\prime })=\frac{\delta ^3\Gamma }{%
i\delta \overline{\psi }(x^{\prime })\delta \psi (y^{\prime })\delta A_\nu
^b(z^{\prime })}\mid _{J=0}  \eqnum{7.7}
\end{equation}
is the quark-gluon proper vertex. 
\begin{equation}
G_1^{ba}(x,y,z)=\int d^4x^{\prime }d^4z^{\prime }S_F(x-x^{\prime })\gamma
^{bc}(x^{\prime },y,z^{\prime })\Delta ^{ca}(z^{\prime }-z)  \eqnum{7.8}
\end{equation}
where $\Delta ^{ca}(z^{\prime }-z)$ is the ghost particle propagator defined
in Eq. (4.7) and 
\begin{equation}
\gamma _1^{bc}(x^{\prime },y,z^{\prime })=\int d^4ud^4v\Delta
^{bd}(y-u)\Gamma ^{cd}(x^{\prime },v,u,z^{\prime })S_F(v-y)  \eqnum{7.9}
\end{equation}
in which 
\begin{equation}
\Gamma ^{cd}(x^{\prime },v,u,z^{\prime })=i\frac{\delta ^4\Gamma }{\delta 
\overline{\psi }(x^{\prime })\delta \psi (v)\delta \overline{C}^c(u)\delta
C^d(z^{\prime })}\mid _{J=0}  \eqnum{7.10}
\end{equation}
is the quark-ghost vertex. Similarly, 
\begin{equation}
G_2^{ba}(x,y,z)=\int d^4y^{\prime }d^4z^{\prime }\gamma ^{bc}(x,y^{\prime
},z^{\prime })S_F(y^{\prime }-y)\Delta ^{ca}(z^{\prime }-z)  \eqnum{7.11}
\end{equation}
where 
\begin{equation}
\gamma _2^{bc}(x,y^{\prime },z^{\prime })=\int d^4ud^4vS_F(x-u)\Delta
^{bd}(x-v)\Gamma ^{dc}(u,y^{\prime },v,z^{\prime }).  \eqnum{7.12}
\end{equation}
On substituting Eqs. (7.5), (7.8) and (7.11) into Eq. (7.1) and then
transform Eq. (7.1) into the momentum space, we have 
\begin{equation}
\begin{array}{c}
S_F(p)\Gamma ^{b\nu }(p,q,k)S_F(q)k^\mu D_{\mu \nu }^{ab}(k) \\ 
=-i\alpha g[S_F(p)\gamma _1^b(p,q,k)-\gamma _2^b(p,q,k)S_F(q)]\Delta
^{ab}(k)]
\end{array}
\eqnum{7.13}
\end{equation}
where we have defined 
\begin{equation}
\begin{array}{c}
\gamma _1^a(p,q,k)=\gamma _1^{ab}(p,q,k)T^b, \\ 
\gamma _2^a(p,q,k)=T^b\gamma _2^{ba}(p,q,k).
\end{array}
\eqnum{7.14}
\end{equation}
Considering that the vertex functions $\Gamma ^{b\nu }(p,q,k)$ and $\gamma
_i^a(p,q,k)$ ($i=1,2$) contain a common delta-function representing the
energy-momentum conservation, we may set

\begin{equation}
\begin{array}{c}
\Gamma ^{a\mu }(p,q,k)=(2\pi )^4\delta ^4(p-q+k)\Lambda ^{a\mu }(p,q,k), \\ 
\gamma _i^a(p,q,k)=(2\pi )^4\delta ^4(p-q+k)\widetilde{\gamma }_i^a(p,q,k)
\end{array}
\eqnum{7.15}
\end{equation}
where $\Lambda ^{a\mu }(p,q,k)$ and $\widetilde{\gamma }_i^a(p,q,k)$ are the
new vertex functions in which $k=q-p$. Noticing the above relations and the
expressions of gluon and ghost particle propagators as given in Eqs. (4.11)
and (4.22), the W-T identity in Eq. (7.13) can be rewritten via the
functions $\Lambda ^{a\mu }(p,q,k)$ and $\widetilde{\gamma }_i^a(p,q,k)$ in
the form 
\begin{eqnarray}
k_\mu \Lambda ^{a\mu }(p,q,k)=ig\chi (k^2)[S_F^{-1}(p)\widetilde{\gamma }%
_2^a(p,q,k)-\widetilde{\gamma }_1^a(p,q,k)S_F^{-1}(q)]  \eqnum{7.16}
\end{eqnarray}
where $\chi (k^2)$ was defined in Eq. (5.13).

Let us turn to discuss the renormalized form of the above W-T identity. It
is well-known that the quark propagator can be expressed in the form 
\begin{equation}
S_F(p)=\frac 1{{\bf p-}m-\Sigma (p)+i\varepsilon }  \eqnum{7.17}
\end{equation}
where ${\bf p=\gamma }^\mu p_\mu $ and $\Sigma (p)$ denotes the quark
self-energy. The above expression can easily be derived from the Dyson
equation [29]. Usually, the quark propagator is renormalized in such a
fashion 
\begin{equation}
S_F(p)=Z_2S_F^R(p)  \eqnum{7.18}
\end{equation}
which implies 
\begin{equation}
\psi (x)=\sqrt{Z_2}\psi _R(x),\text{ }\overline{\psi }(x)=\sqrt{Z_2}%
\overline{\psi }_R(x).  \eqnum{7.19}
\end{equation}
From the relations in Eqs. (7.19) and (5.18), it is clearly seen that the
vertex defined in Eq. (7.7) is renormalized as 
\begin{equation}
\Gamma ^{a\mu }(x,y,z)=Z_2^{-1}Z_3^{-\frac 12}\Gamma _R^{a\mu }(x,y,z) 
\eqnum{7.20}
\end{equation}
which leads to 
\begin{equation}
\Lambda ^{a\mu }(p,q,k)=Z_2^{-1}Z_3^{-\frac 12}\Lambda _R^{a\mu }(p,q,k) 
\eqnum{7.21}
\end{equation}
The functions $\widetilde{\gamma }_i^a(p,q,k)$ are, in general, divergent in
the perturbative calculation. These functions are assumed to be renormalized
in such a manner 
\begin{equation}
\widetilde{\gamma }_i^a(p,q,k)=Z_\gamma ^{-1}\widetilde{\gamma }%
_{iR}^a(p,q,k).  \eqnum{7.22}
\end{equation}
where $Z_\gamma $ is the renormalization constant of the functions $%
\widetilde{\gamma }_i^a(p,q,k)$. Based on the relations in Eqs. (5.22),
(7.18), (7.21), and (7.22), Eq. (7.16) can be represented in terms of the
renormalized quantities 
\begin{eqnarray}
k_\mu \Lambda _R^{a\mu }(p,q,k)=ig_R\chi _R(k^2)[S_F^{R-1}(p)\widetilde{%
\gamma }_{2R}^a(p,q,k)-\widetilde{\gamma }_{1R}^a(p,q,k)S_F^{R-1}(q)] 
\eqnum{7.23}
\end{eqnarray}
where $g_R$ is the renormalized coupling constant defined by 
\begin{equation}
g_R=\widetilde{Z_3}Z_3^{\frac 12}\widetilde{Z}_\gamma ^{-1}g  \eqnum{7.24}
\end{equation}
It is well-known that 
\begin{equation}
g_R=\widetilde{Z_3}Z_3^{\frac 12}\widetilde{Z}_1^{-1}g  \eqnum{7.25}
\end{equation}
where $\widetilde{Z}_1$ is the ghost vertex renormalization constant as
defined in Eq. (5.24). The relation in Eq. (7.25) ordinarily is determined
from the renormalization of S-matrix elements. In comparison of Eq. (7.24)
with Eq. (7.25), we see 
\begin{equation}
\widetilde{Z}_\gamma =\widetilde{Z}_1  \eqnum{7.26}
\end{equation}
which means that the functions $\widetilde{\gamma }_i^a(p,q,k)$ are
renormalized in the same way as for the ghost vertex.

In the conventional discussion of the vertex renormalization, one considers
such a vertex denoted by $\widetilde{\Lambda }^{a\mu }(p,q,k)$ that it is
defined from $\Lambda ^{a\mu }(p,q,k)$ by taking out a coupling constant.
Obviously, the W-T identity obeyed by the $\widetilde{\Lambda }^{a\mu
}(p,q,k)$ can be written out from (7.16) by taking away the coupling
constant on the RHS of Eq. (7.16), that is 
\begin{eqnarray}
k_\mu \widetilde{\Lambda }^{a\mu }(p,q,k)=i\chi (k^2)[S_F^{-1}(p)\widetilde{%
\gamma }_2^a(p,q,k)-\widetilde{\gamma }_1^a(p,q,k)S_F^{-1}(q)].  \eqnum{7.27}
\end{eqnarray}
The renormalization of the vertex $\widetilde{\Lambda }^{a\mu }(p,q,k)$
usually is defined by 
\begin{equation}
\widetilde{\Lambda }^{a\mu }(p,q,k)=Z_F^{-1}\widetilde{\Lambda }_R^{a\mu
}(p,q,k)  \eqnum{7.28}
\end{equation}
where $Z_F$ is the quark-gluon vertex renormalization constant. When Eqs.
(5.22), (7.18), (7.22) and (7.28) are inserted into Eq. (7.27) and then
multiplying the both sides of Eq. (7.27) with a renormalized coupling
constant, we arrive at 
\begin{eqnarray}
k_\mu \Lambda _R^{a\mu }(p,q,k)=iZ_F\widetilde{Z}_3Z_2^{-1}Z_\gamma
^{-1}g_R\chi _R(k^2)[S_F^{R-1}(p)\widetilde{\gamma }_{2R}^a(p,q,k)-%
\widetilde{\gamma }_{1R}^a(p,q,k)S_F^{R-1}(q)]  \eqnum{7.29}
\end{eqnarray}
where 
\begin{equation}
\Lambda _R^{a\mu }(p,q,k)=g_R\widetilde{\Lambda }_R^{a\mu }(p,q,k). 
\eqnum{7.30}
\end{equation}
In comparison of Eq. (7.29) with Eq. (7.23) and considering the equality in
Eq. (7.26), we find, the following identity must hold 
\begin{equation}
\frac{Z_F}{Z_2}=\frac{\widetilde{Z}_1}{\widetilde{Z}_3}.  \eqnum{7.31}
\end{equation}
Combining the relations in Eqs. (5.28), (6.24) and (7.31), we have 
\begin{equation}
\frac{Z_F}{Z_2}=\frac{\widetilde{Z}_1}{\widetilde{Z}_3}=\frac{Z_1}{Z_3}=%
\frac{Z_4}{Z_1}.  \eqnum{7.32}
\end{equation}
This just is the well-known S-T identity. This identity was obtained from
the massless QCD and now, as has just been proved, it also holds for the
massive QCD.

\section{ Effective coupling constant and gluon mass}

This section and the next section are used to perform one-loop
renormalization of the massive QCD by using the renormalization group
approach. As argued in our previous paper [30-32], when the renormalization
is carried out in the mass-dependent momentum space subtraction scheme, the
solutions to the RGEs satisfied by renormalized wave functions, propagators
and vertices can be uniquely determined by the boundary conditions of the
renormalized wave functions, propagators and vertices. In this case, an
exact S-matrix element can be written in the form as given in the
tree-diagram approximation provided that the coupling constant and particle
masses in the matrix element are replaced by their effective (running) ones
which are given by solving their renormalization group equations. Therefore,
the task of renormalization is reduced to find the solutions of the RGEs for
the renormalized coupling constant and particle masses. Suppose $F_R$ is a
renormalized quantity. In the multiplicative renormalization, it is related
to the unrenormalized one $F$ in such a way 
\begin{eqnarray}
F=Z_FF_R  \eqnum{8.1}
\end{eqnarray}
where $Z_F$ is the renormalization constant of $F$. The $Z_F$ and $F_R$ are
all functions of the renormalization point $\mu =\mu _0e^t$ where $\mu _0$
is a fixed renormalization point corresponding the zero value of the group
parameter $t$. Differentiating Eq. (8.1) with respect to $\mu $ and noticing
that the $F$ is independent of $\mu $, we immediately obtain a
renormalization group equation (RGE) satisfied by the function $F_R$ [21-23] 
\begin{eqnarray}
\mu \frac{dF_R}{d\mu }+\gamma _FF_R=0  \eqnum{8.2}
\end{eqnarray}
where $\gamma _F$ is the anomalous dimension defined by 
\begin{eqnarray}
\gamma _F=\mu \frac d{d\mu }\ln Z_F.  \eqnum{8.3}
\end{eqnarray}
Since the renormalization constant is dimensionless, the anomalous dimension
can only depend on the ratio ${\beta =\frac{m_R}\mu }${\ where }$m_R$
denotes a, renormalized mass and ${\gamma }_F{=\gamma }_F{(g}_R{,\beta )}$
in which $g_R$ is the renormalized coupling constant and depends on $\mu $.
Since the renormalization point is a momentum taken to subtract the
divergence, we may set $\mu =\mu _0\lambda $ where $\lambda =e^t$ which will
be taken to be the same as in the scaling transformation of momentum $%
p=p_0\lambda $. In the above, $\mu _0$ and $p_0$ are the fixed
renormalization point and momentum respectively. When we set $F$ to be the
coupling constant $g$ and noticing $\mu \frac d{d\mu }=\lambda \frac d{%
d\lambda }$, one can write from Eq. (8.2) the RGE for the renormalized
coupling constant 
\begin{equation}
\lambda \frac{dg_R(\lambda )}{d\lambda }+\gamma _g(\lambda )g_R(\lambda )=0 
\eqnum{8.4}
\end{equation}
with 
\begin{equation}
\gamma _g=\mu \frac d{d\mu }\ln Z_g.  \eqnum{8.5}
\end{equation}
According to the definition in Eq. (8.1) and the relation in Eq. (7.25), we
may take, 
\begin{equation}
Z_g=\frac{\widetilde{Z}_1}{\tilde Z_3Z_3^{\frac 12}}  \eqnum{8.6}
\end{equation}
to calculate the anomalous dimension. As denoted in Eqs. (4.25) and (5.24),
the renormalization constants $Z_3$, $\tilde Z_3$ and $\tilde Z_1$ are
determined by the gluon self-energy, the ghost article self-energy and the
ghost vertex correction, respectively. At one-loop level, the gluon
self-energy is depicted in Figs. (1a)-(1d), the ghost article self-energy is
shown in Fig. (2) and the ghost vertex correction is represented in Figs.
(3a) and (3b). According to the Feynman rules which are the same as those
for the massless QCD [16] except that the gluon propagator and the ghost
particle one are now given in Eqs. (4.14) and (4.10), the expressions of the
self-energies and the vertex correction are easily written out. For the
gluon one-loop self-energy denoted by $-i\Pi _{\mu \nu }^{ab}(k)$, one can
write 
\begin{equation}
\Pi _{\mu \nu }^{ab}(k)=\sum_{i=1}^4\Pi _{\mu \nu }^{(i)ab}(k)  \eqnum{8.7}
\end{equation}
where $\Pi _{\mu \nu )}^{(1)ab}(k)$-$\Pi _{\mu \nu )}^{(4)ab}(k)$ represent
the self-energies given in turn by Figs.(1a)-(1d). They are separately
represented in the following: 
\begin{equation}
\begin{array}{c}
\Pi _{\mu \nu }^{(1)ab}(k)=i\delta ^{ab}\frac 32g^2\int \frac{d^4l}{(2\pi )^4%
}\frac{g^{\lambda \lambda ^{\prime }}g^{\rho \rho ^{\prime }}}{%
[l^2-M^2+i\varepsilon ][(l+k)^2-M^2+i\varepsilon ]}[g_{\mu \lambda
}(l+2k)_\rho -g_{\lambda \rho }(2l+k)_\mu \\ 
+g_{\rho \mu }(l-k)_\lambda ][g_{\nu \rho ^{\prime }}(l-k)_{\lambda ^{\prime
}}-g_{\lambda ^{\prime }\rho ^{\prime }}(2l+k)_\nu +g_{\lambda ^{^{\prime
}}\nu }(l+2k)_{\rho ^{\prime }}],
\end{array}
\eqnum{8.8}
\end{equation}
\begin{equation}
\Pi _{\mu \nu }^{(2)ab}(k)=-i\delta ^{ab}3g^2\int \frac{d^4l}{(2\pi )^4}%
\frac{(l+k)_\mu l_\nu }{[(l+k)^2-M^2+i\varepsilon ][l^2-M^2+i\varepsilon ]},
\eqnum{8.9}
\end{equation}
\begin{equation}
\Pi _{\mu \nu }^{(3)ab}(k)=-i\delta ^{ab}3g^2\int \frac{d^4l}{(2\pi )^4}%
\frac{g^{\lambda \rho }}{(l^2-M^2+i\varepsilon )}(g_{\mu \nu }g_{\lambda
\rho }-g_{\mu \rho }g_{\lambda \nu })  \eqnum{8.10}
\end{equation}
and 
\begin{equation}
\begin{array}{c}
\Pi _{\mu \nu }^{(4)ab}(k)=-i\delta ^{ab}\frac 12g^2\int \frac{d^4l}{(2\pi
)^4}\frac 1{[(l-k)^2-m^2+i\varepsilon ][l^2-m^2+i\varepsilon ]} \\ 
\times Tr[\gamma _\mu ({\bf l}-{\bf k}+m)\gamma _\nu ({\bf l}+m)]
\end{array}
\eqnum{8.11}
\end{equation}
where ${\bf l=}\gamma ^\lambda l_\lambda $, ${\bf k=}\gamma ^\lambda
k_\lambda $. In the above, $f^{acd}f^{bcd}=3\delta ^{ab}$ and $Tr(T^aT^b)=%
\frac 12\delta ^{ab}$ have been considered. It should be noted that in
writing Eqs. (8.8)-(8.10), we choose to work in the Feynman gauge for
simplicity. This choice is based on the fact that the massive QCD has been
proved to be an unitary theory [18], that is to say, the S-matrix elements
evaluated from the massive QCD are independent of gauge parameter.
Therefore, we are allowed to choose a convenient gauge in the calculation.
From Eqs. (8.8)-(8.11), it is clearly seen that 
\begin{equation}
\Pi _{\mu \nu }^{ab}(k)=\delta ^{ab}\Pi _{\mu \nu }(k)=\delta
^{ab}\sum_{i=1}^4\Pi _{\mu \nu }^{(i)}(k).  \eqnum{8.12}
\end{equation}
By the dimensional regularization approach [33-37], the divergent integrals
over $l$ in Eqs. (8.8)-(8.11) can be regularized in a $n$-dimensional space
and easily calculated. The results are 
\begin{equation}
\begin{array}{c}
\Pi _{\mu \nu }^{(1)}(k)=-\frac 32\frac{g^2}{(4\pi )^2}\int_0^1dx\frac 1{%
\varepsilon [k^2x(x-1)+M^2]^\varepsilon }\{g_{\mu \nu }[11x(x-1) \\ 
+5)k^2+9M^2]+2[5x(x-1)-1]k_\mu k_\nu \},
\end{array}
\eqnum{8.13}
\end{equation}
\begin{equation}
\begin{array}{c}
\Pi _{\mu \nu }^{(2)}(k)=\frac 32\frac{g^2}{(4\pi )^2}\int_0^1dx\frac 1{%
\varepsilon [k^2x(x-1)+M^2]^\varepsilon }\{[k^2x(x-1) \\ 
+M^2]g_{\mu \nu }+2x(x-1)k_\mu k_\nu \},
\end{array}
\eqnum{8.14}
\end{equation}
\begin{equation}
\Pi _{\mu \nu }^{(3)}(k)=\frac{9g^2}{(4\pi )^2}\frac{M^2}\varepsilon g_{\mu
\nu }  \eqnum{8.15}
\end{equation}
and 
\begin{equation}
\Pi _{\mu \nu }^{(4)}(k)=-\frac{4g^2}{(4\pi )^2}\int_0^1dx\frac{k^2x(x-1)}{%
\varepsilon [k^2x(x-1)+m^2]^\varepsilon }[g_{\mu \nu }-\frac{k_\mu k_\nu }{%
k^2}]  \eqnum{8.16}
\end{equation}
where $\varepsilon =2-\frac n2\rightarrow 0$ when $n\rightarrow 4$. In Eqs.
(8.13)-(8.16), except for the $\varepsilon $ in the factor $1/\varepsilon
[k^2x(x-1)+M^2]^\varepsilon $ and $1/\varepsilon [k^2x(x-1)+m^2]^\varepsilon 
$, we have set $\varepsilon \rightarrow 0$ in the other factors and terms by
the consideration that this operation does not affect the calculated result
of the anomalous dimension. According to the decomposition shown in Eqs.
(4.15) and (4.16) and noticing $g_{\mu \nu }={\cal P}_T^{\mu \nu }+{\cal P}%
_L^{\mu \nu }$, it is easy to get the transverse part of $\Pi _{\mu \nu }(k)$
from Eqs. (8.13)-(8.16) and furthermore, based on the decomposition denoted
in Eq. (4.20), the functions $\Pi _1(k^2)$ and $\Pi _2(k^2)$ can be written
out. The results are 
\begin{equation}
\Pi _1(k^2)=-\frac{g^2}{(4\pi )^2}\int_0^1dx\{\frac{15[2x(x-1)+1]}{%
2\varepsilon [k^2x(x-1)+M^2]^\varepsilon }+\sum\limits_{i=1}^{N_f}\frac{%
4x(x-1)}{\varepsilon [k^2x(x-1)+m_i^2]^\varepsilon }\}  \eqnum{8.17}
\end{equation}
and 
\begin{equation}
\Pi _2(k^2)=-\frac{g^2}{(4\pi )^2}\{\int_0^1dx\frac{12}{\varepsilon
[k^2x(x-1)+M^2]^\varepsilon }-\frac{27}{2\varepsilon M^{2\varepsilon }}\}. 
\eqnum{8.18}
\end{equation}
It is clear that the both functions $\Pi _1(k^2)$ and $\Pi _2(k^2)$ are
divergent in the four-dimensional space-time. When the divergences are
subtracted in the mass-dependent momentum space subtraction scheme [34-37],
in accordance with the definition in Eq. (4.25), we immediately obtain from
the expression in Eq. (8.17) the one-loop renormalization constant $Z_3$ as
follows 
\begin{equation}
\begin{array}{c}
Z_3=1-\Pi _1(\mu ^2) \\ 
=1+\frac{g^2}{(4\pi )^2}\int_0^1dx\{\frac{15[2x(x-1)+1]}{2\varepsilon [\mu
^2x(x-1)+M^2]^\varepsilon }+\sum\limits_{i=1}^{N_f}\frac{4x(x-1)}{%
\varepsilon [\mu ^2x(x-1)+m_i^2]^\varepsilon }\}.
\end{array}
\eqnum{8.19}
\end{equation}

Next, we turn to the ghost particle one-loop self-energy denoted by $%
-i\Omega ^{ab}(q)$. From Fig. (2), in Feynman gauge, one can write 
\begin{equation}
\Omega ^{ab}(q)=i\delta ^{ab}3g^2\int \frac{d^4l}{(2\pi )^4}\frac{q\cdot
(q-l)}{[(q-l)^2-M^2+i\varepsilon ][l^2-M^2+i\varepsilon ]}.  \eqnum{8.20}
\end{equation}
By the dimensional regularization, it is easy to get 
\begin{equation}
\Omega ^{ab}(q)=\delta ^{ab}q^2\hat \Omega (q^2)  \eqnum{8.21}
\end{equation}
where 
\begin{equation}
\hat \Omega (q^2)=\frac{g^2}{(4\pi )^2}\int_0^1dx\frac{3(x-1)}{\varepsilon
[q^2x(x-1)+M^2]^\varepsilon }.  \eqnum{8.22}
\end{equation}
According to the definition given in Eq. (4.25) and the above expression ,
the one-loop renormalization constant of ghost particle propagator is of the
form 
\begin{equation}
\widetilde{Z}_3=1-\hat \Omega (\mu ^2)=1-\frac{g^2}{(4\pi )^2}\int_0^1dx%
\frac{3(x-1)}{\varepsilon [\mu ^2x(x-1)+M^2]^\varepsilon }.  \eqnum{8.23}
\end{equation}

Now, let us discuss the ghost vertex renormalization. In the one-loop
approximation. the vertex defined by extracting out a coupling constant is
expressed as 
\begin{equation}
\widetilde{\Lambda }_\lambda ^{abc}(p,q)=f^{abc}p_\lambda +\Lambda
_{1\lambda }^{abc}(p,q)+\Lambda _{2\lambda }^{abc}(p,q)  \eqnum{8.24}
\end{equation}
where the first term is the bare vertex, the second and the third terms
stand for the one-loop vertex corrections shown in Figs. (3a) and (3b)
respectively. In the Feynman gauge, the vertex corrections are expressed as 
\begin{equation}
\Lambda _{1\lambda }^{abc}(p,q)=-if^{abc}\frac 32g^2\int \frac{d^4l}{(2\pi
)^4}\frac{p\cdot (q-l)(p-l)_\lambda }{[l^2-M^2+i\varepsilon
][(p-l)^2-M^2+i\varepsilon ][(q-l)^2-M^2+i\varepsilon ]}  \eqnum{8.25}
\end{equation}
and 
\begin{equation}
\Lambda _{2\lambda }^{abc}(p,q)=if^{abc}\frac 32g^2\int \frac{d^4l}{(2\pi )^4%
}\frac{l\cdot (p-q-l)p_\lambda -p\cdot lq_\lambda +p\cdot (2q-p+l)l_\lambda 
}{[l^2-M^2+i\varepsilon ][(p-l)^2-M^2+i\varepsilon
][(q-l)^2-M^2+i\varepsilon ]}  \eqnum{8.26}
\end{equation}
where $f^{acd}f^{ebf}f^{dfc}=-\frac 32f^{abc}$ has been noted. By employing
the dimensional regularization to compute the above integrals, it is not
difficult to get 
\begin{equation}
\Lambda _{1\lambda }^{abc}(p,q)=f^{abc}\frac 32\frac{g^2}{(4\pi )^2}%
\int_0^1dx\int_0^1dy\{\frac{\frac 12yp_\lambda }{\varepsilon \Theta
_{xy}^\varepsilon }-\frac 1{\Theta _{xy}}[p_\lambda A_1(p,q)+q_\lambda
B_1(p,q)]-\frac 18p_\lambda \}  \eqnum{8.27}
\end{equation}
where 
\begin{equation}
\begin{array}{c}
\Theta _{xy}=p^2xy(xy-1)+q^2[(x-1)^2y+(x-1)]y-2p\cdot qx(x-1)y^2+M^2, \\ 
A_1(p,q)=\{p\cdot q[1+(x-1)y]-p^2xy\}(1-xy)y, \\ 
B_1(p,q)=\{p\cdot q[1+(x-1)y]-p^2xy\}(x-1)y^2
\end{array}
\eqnum{8.28}
\end{equation}
and 
\begin{equation}
\Lambda _{2\lambda }^{abc}(p,q)=f^{abc}\frac 32\frac{g^2}{(4\pi )^2}%
\int_0^1dx\int_0^1dy\{\frac{\frac 32yp_\lambda }{\varepsilon \Theta
_{xy}^\varepsilon }+\frac 1{\Theta _{xy}}[p_\lambda A_2(p,q)+q_\lambda
B_2(p,q)]-\frac 38p_\lambda \}  \eqnum{8.29}
\end{equation}
where 
\begin{equation}
\begin{array}{c}
A_2(p,q)=\{p^2(2xy-x^2y^2-1)-q^2[(x-1)y-1](x-1)y \\ 
+p\cdot q[2-(3x-2)y+2x(x-1)y^2]\}y, \\ 
B_2(p,q)=[p\cdot q(x-1)-p^2x]y^2.
\end{array}
\eqnum{8.30}
\end{equation}
The divergences in the both vertices $\Lambda _{1\lambda }^{abc}(p,q)$ and $%
\Lambda _{2\lambda }^{abc}(p,q)$ may be subtracted at the renormalization
point $p^2=q^2=\mu ^2$ which implies $k=p-q=0$, being consistent with the
momentum conservation held at the vertices. Upon substituting Eqs. (8.27)
and (8.29) in Eq. (8.24), at the renormalization point, one can get 
\begin{equation}
\widetilde{\Lambda }_\lambda ^{abc}(p,q)\mid _{p^2=q^2=\mu
^2}=f^{abc}p_\lambda (1+\widetilde{L}_1)=\widetilde{Z}_1^{-1}f^{abc}p_\lambda
\eqnum{8.31}
\end{equation}
where 
\begin{equation}
\widetilde{Z}_1=1-\widetilde{L}_1=1-\frac{3g^2}{(4\pi )^2}\int_0^1dx\{\frac x%
{\varepsilon [\mu ^2x(x-1)+M^2]^\varepsilon }-\frac{x^2(x-1)\mu ^2}{\mu
^2x(x-1)+M^2}-\frac 14\}  \eqnum{8.32}
\end{equation}
which is the one-loop renormalization constant of the ghost vertex.

Now we are ready to calculate the anomalous dimension $\gamma _g(\lambda )$.
Substituting the expressions in Eqs. (8.6), (8.19), (8.23) and (8.32) into
Eq. (8.5), it is easy to find an analytical expression of the anomalous
dimension $\gamma _g(\lambda )$. When we set $\frac M\mu =\frac \beta \lambda
$ and $\frac{m_i}\mu =\frac{\rho _i}\lambda $ with defining $\beta =\frac M%
\Lambda $ and $\rho _i=\frac{m_i}\Lambda $ (here we have set $\mu _0\equiv
\Lambda $), the expression of $\gamma _g(\lambda )$, in the approximation of
order $g^2$, is given by 
\begin{equation}
\gamma _g(\lambda )=\lim\limits_{\varepsilon \rightarrow 0}[\mu \frac d{d\mu 
}\ln \widetilde{Z}_1-\mu \frac d{d\mu }\ln \widetilde{Z}_3-\frac 12\mu \frac 
d{d\mu }\ln Z_3]=\frac{g_R^2}{(4\pi )^2}F(\lambda )  \eqnum{8.33}
\end{equation}
where 
\begin{equation}
\begin{array}{c}
F(\lambda )=\frac{19}2-\frac{15\beta ^2}{\lambda ^2}+\frac{3\lambda ^2}{%
2(\lambda ^2-4\beta ^2)}-(8-\frac{10\beta ^2}{\lambda ^2}-\frac{\lambda ^2}{%
\lambda ^2-4\beta ^2})\frac{3\beta ^2}{\lambda \sqrt{\lambda ^2-4\beta ^2}}
\\ 
\times \ln \frac{\lambda -\sqrt{\lambda ^2-4\beta ^2}}{\lambda +\sqrt{%
\lambda ^2-4\beta ^2}}-\frac 23\sum\limits_{i=1}^{N_f}[1+\frac{6\rho _i^2}{%
\lambda ^2}-\frac{12\rho _i^4}{\lambda ^3\sqrt{\lambda ^2-4\rho _i^2}}\ln 
\frac{\lambda -\sqrt{\lambda ^2-4\rho _i^2}}{\lambda +\sqrt{\lambda ^2-4\rho
_i^2}}]
\end{array}
\eqnum{8.34}
\end{equation}
in which $N_f$ denotes the number of quark flavors. We would like to note
that the fixed renormalization point $\Lambda $ in $\beta $ and $\rho _i$
can be taken arbitrarily. For example, the $\Lambda $ may be chosen to be
the mass of the quark of $N_f$ -th flavor. In this case, $\beta =M/m_{N_f}$
and $\rho _i=m_i/m_{N_f}$. In practice, the $\Lambda $ will be treated as a
scaling parameter of renormalization.

With the $\gamma _g(\lambda )$ given above, the equation in Eq. (8.4) can be
solved to give the effective coupling constant as follows 
\begin{equation}
\alpha _R(\lambda )=\frac{\alpha _R}{1+\frac{\alpha _R}{2\pi }G(\lambda )} 
\eqnum{8.35}
\end{equation}
where $\alpha _R(\lambda )=g_R^2(\lambda )/4\pi $, $\alpha _R=\alpha _R(1)$
and 
\begin{equation}
G(\lambda )=\int_1^\lambda \frac{d\lambda }\lambda F(\lambda )=\varphi
_1(\lambda )-\varphi _1(1)-\frac 13\sum\limits_{i=1}^{N_f}[\varphi
_2^i(\lambda )-\varphi _2^i(1)]  \eqnum{8.36}
\end{equation}
in which 
\begin{equation}
\varphi _1(\lambda )=[(19-\frac{10\beta ^2}{\lambda ^2})\frac{\sqrt{\lambda
^2-4\beta ^2}}{4\lambda }+\frac{3\lambda }{4\sqrt{\lambda ^2-4\beta ^2}}]\ln 
\frac{\lambda +\sqrt{\lambda ^2-4\beta ^2}}{\lambda -\sqrt{\lambda ^2-4\beta
^2}}+\frac{5\beta ^2}{\lambda ^2},  \eqnum{8.37}
\end{equation}
\begin{equation}
\varphi _2^i(\lambda )=(1+\frac{2\rho _i^2}{\lambda ^2})\frac{\sqrt{\lambda
^2-4\rho _i^2}}\lambda \ln \frac{\lambda +\sqrt{\lambda ^2-4\rho _i^2}}{%
\lambda -\sqrt{\lambda ^2-4\rho _i^2}}-\frac{4\rho _i^2}{\lambda ^2} 
\eqnum{8.38}
\end{equation}
and $\varphi _1(1)=\varphi _1(\lambda )\mid _{\lambda =1}$, $\varphi
_2^i(1)=\varphi _2^i(\lambda )\mid _{_{\lambda =1}}$. In the large momentum
limit ($\lambda \rightarrow \infty $), we have 
\begin{equation}
G(\lambda )=(11-\frac 23N_f)\ln \lambda .  \eqnum{8.39}
\end{equation}
This just is the result for massless QCD which was obtained previously in
the minimal subtraction scheme [38-40]. It should be noted that the
expressions in Eqs. (8.34), (8.37) and (8.38) are obtained at the timelike
subtraction point where the $\lambda $ is a real variable. We may also take
spacelike momentum subtraction. For this kind of subtraction, corresponding
to $\mu \rightarrow i\mu $, the variable $\lambda $ in Eqs. (8.34), (8.37)
and (8.38) should be replaced by $i\lambda $ where $\lambda $ is still a
real variable. It is easy to see that the function in Eq. (8.39) is the same
for the both subtractions.

The behavior of the function $\alpha _R(\lambda )$ is graphically described
in Figs. (4)-(6). Figs. (4) and (5) represent respectively the effective
coupling constants obtained at the timelike subtraction point and the
spacelike subtraction point, where we take the flavor $N_f=3$ as an
illustration. For comparison, we show in each of the figures three effective
coupling constants which are obtained by the massive QCD, the massless QCD
and the minimal subtraction, respectively. These effective coupling
constants are respectively represented by the solid, dashed and dotted lines
in the figures. To exhibit the dependence of the effective coupling constant
on the number of quark flavors, in Fig. (6), we show three effective
coupling constants given by the timelike momentum subtraction. The solid,
dashed and dotted lines in the figure represent the effective coupling
constants obtained by taking $N_f=2,3$ and $4$, respectively. In our test,
we find that the behavior of the $\alpha _R(\lambda )$ sensitively depends
on the choice of the constant $\alpha _R$, the scaling parameter $\Lambda $,
gluon mass $M_R$ and the quark masses $m_R$. Certainly, the parameters $%
\alpha _R,$ $\Lambda $, $M_R$ and $m_R$ should be determined by fitting to
experimental data. In our calculation, as an illustration, we take the quark
masses to be constituent quark ones. The masses of up, down, strange and
charm quarks are taken to be $m_u=m_d=350MeV$, $m_s=500MeV$ and $m_c=1500MeV$%
. For the other parameters, we take $\alpha _R=0.2$, $M_R=600MeV$, $\Lambda
=500MeV$ in Figs. (4) and (5) and $\Lambda =1500MeV$ in Fig. (6). From Fig.
(4) it is seen that the effective coupling constant given by the massive QCD
is an analytical function with a maximum at $\lambda =1.346$, the effective
coupling constant given by the massless QCD is also an analytical function
with a peak around $\lambda =1$, but the effective coupling constant given
by the minimal subtraction has a singularity at $\lambda =$ $0.1746$. Fig.
(5) indicates that the effective coupling constant given by the massive QCD
(where gluon mass $M_R=600MeV$), similar to the one given by the minimal
subtraction, has a Landau pole at $\lambda =0.1845$ which implies that the
coupling constant is not applicable in the region $\lambda \leq 0.1845$.
However, if the gluon mass is taken to be $M_R\leq 425.75MeV$, we find, the
Landau pole disappears and the effective coupling constant, analogous to the
effective coupling constant given by the massless QCD, becomes a smooth
function in the whole region of momentum as illustrated by the dotted-dashed
line in Fig. (5) which represents the effective coupling constant given by
taking $M_R=425.75MeV$. From Fig. (6) it is clear to see that in the low and
intermediate region of momentum, the larger the flavor number $N_f$, the
smaller the maximum of the effective coupling constant is and the positions
of the maxima for different coupling constants are different from one
another. This property of the effective coupling constants would give a
notable effect on the theoretical hadron spectrum because for the
calculation of hadron spectrum, the quarks in a hadron are assumed to move
not too fast as suggested in the nonperturbative quark potential model,
therefore, the behavior of the effective coupling constant in the low and
intermediate domain of energy would play a dominate role In this case.

Let us proceed to derive the one-loop effective gluon mass. Setting $F_R=M_R$
in Eq. (8.2), we have the RGE for the renormalized gluon mass 
\begin{equation}
\lambda \frac{dM_R(\lambda )}{d\lambda }+\gamma _M(\lambda )M_R(\lambda )=0 
\eqnum{8.40}
\end{equation}
where 
\begin{equation}
\gamma _M(\lambda )=\mu \frac d{d\mu }\ln Z_M.  \eqnum{8.41}
\end{equation}
From the last equality in Eq. (4.25) and Eqs. (8.17) and (8.18), in the
approximation of order $g^2$, we can write 
\begin{equation}
\begin{array}{c}
Z_M=1+\frac 12[\Pi _1(\mu ^2)+\Pi _2(\mu ^2)] \\ 
=1-\frac{g^2}{(4\pi )^2}\{\int_0^1dx[\frac{3[10x(x-1)+13]}{4\varepsilon
[k^2x(x-1)+M^2]^\varepsilon }+\sum\limits_{i=1}^{N_f}\frac{2x(x-1)}{%
\varepsilon [k^2x(x-1)+m_i^2]^\varepsilon }]-\frac{27}{4\varepsilon
M^{2\varepsilon }}\}.
\end{array}
\eqnum{8.42}
\end{equation}
On inserting Eq. (8.42) into Eq. (8.41) and completing the differentiation
with respect to $\mu $ and the integration over $x$, we find 
\begin{equation}
\begin{array}{c}
\gamma _M(\lambda )=\frac{g_R^2}{(4\pi )^2}\{17-\frac{15\beta ^2}{\lambda ^2}%
-\frac{3\beta ^2(13\lambda ^2-10\beta ^2)}{\lambda ^3\sqrt{\lambda ^2-4\beta
^2}}\ln \frac{\lambda -\sqrt{\lambda ^2-4\beta ^2}}{\lambda +\sqrt{\lambda
^2-4\beta ^2}} \\ 
-\frac 23\sum\limits_{i=1}^{N_f}[1+\frac{6\rho _i^2}{\lambda ^2}-\frac{%
12\rho _i^4}{\lambda ^3\sqrt{\lambda ^2-4\rho _i^2}}\ln \frac{\lambda -\sqrt{%
\lambda ^2-4\rho _i^2}}{\lambda +\sqrt{\lambda ^2-4\rho _i^2}}]\}.
\end{array}
\eqnum{8.43}
\end{equation}
With this anomalous dimension, the RGE in Eq. (8.40) can be solved to give
an effective gluon mass such that 
\begin{equation}
M_R(\lambda )=M_Re^{-S_g(\lambda )}  \eqnum{8.44}
\end{equation}
where $M_R=M_R(1)$ and 
\begin{equation}
S_g(\lambda )=\int_1^\lambda \frac{d\lambda }\lambda \gamma _M(\lambda ). 
\eqnum{8.45}
\end{equation}
In general, the coupling constant $g_R$ in Eq. (8.43) may be taken to be the
effective one. If the coupling constant is taken to be the constant $g_R$,
the function $S_g(\lambda )$ can be explicitly represented as 
\begin{equation}
S_g(\lambda )=\frac{\alpha _R}{4\pi }\{\varphi _3(\lambda )-\varphi _3(1)-%
\frac 13\sum\limits_{i=1}^{N_f}[\varphi _2^i(\lambda )-\varphi _2^i(1)]\} 
\eqnum{8.46}
\end{equation}
where 
\begin{equation}
\varphi _3(\lambda )=(17-\frac{5\beta ^2}{\lambda ^2})\frac{\sqrt{\lambda
^2-4\beta ^2}}{2\lambda }\ln \frac{\lambda +\sqrt{\lambda ^2-4\beta ^2}}{%
\lambda -\sqrt{\lambda ^2-4\beta ^2}}+\frac{5\beta ^2}{\lambda ^2} 
\eqnum{8.47}
\end{equation}
and $\varphi _2^i(\lambda )$ was given in Eq. (8.38). In the large momentum
limit, 
\begin{equation}
S_g(\lambda )=\frac{\alpha _R}{4\pi }(17-\frac 23N_f)\ln \lambda . 
\eqnum{8.48}
\end{equation}
Therefore, we have 
\begin{equation}
\lim\limits_{\lambda \rightarrow \infty }M_R(\lambda )=0  \eqnum{8.49}
\end{equation}
which exhibits the asymptotically free behavior.

The behavior of the effective gluon mass $M_R(\lambda )$ may be discussed in
the way similar to the discussion of the effective coupling constant. Here
we only limit ourself to show in Fig. (7) the behavior of the effective
gluon mass written in Eqs. (8.44)-(9.46) with taking $N_f=3$ and the
coupling constant being a constant. In Fig. (7), the solid line and the
dashed line represent the effective gluon masses given by the timelike
momentum subtraction and spacelike momentum subtraction, respectively. The
solid line exhibits that for the timelike momentum, the effective gluon mass
is an analytical function with a maximum at $\lambda =1.233$. When $\lambda $
goes to infinity, the $M_R(\lambda )$ tends to zero rather rapidly, while,
when $\lambda $ goes to zero, the $M_R(\lambda )$ abruptly falls to zero.
The dashed line tells us that for the spacelike momentum, the $M_R(\lambda )$
keeps a constant $M_R$ in the region [0,1] of $\lambda $, while when $%
\lambda \rightarrow \infty $, the $M_R(\lambda )$ smoothly tends to zero.

\section{Effective quark mass}

Before deriving the one-loop effective quark mass, we need first to discuss
the subtraction of the quark one-loop self-energy on the basis of the W-T
identity represented in Eq. (7.16). For later convenience, the identity in
Eq. (7.16) will be given in another form. Introducing new vertex functions $%
\hat \Lambda ^{a\mu }(p,q)$ and $\hat \gamma _i^a(p,q)$ defined by 
\begin{equation}
\begin{array}{c}
\Gamma ^{a\mu }(p,q,k)=(2\pi )^4\delta ^4(p-q+k)ig\hat \Lambda ^{a\mu }(p,q)
\\ 
\gamma _i^a(p,q,k)=-(2\pi )^4\delta ^4(p-q+k)\hat \gamma _i^a(p,q)
\end{array}
\eqnum{9.1}
\end{equation}
where $i=1,2$ and $\hat \gamma _i^a(p,q)=-\widetilde{\gamma }_i^a(p,q)$ and
considering $k=q-p$, Eq. (7.16) can be rewritten as 
\begin{eqnarray}
(p-q)_\mu \hat \Lambda ^{a\mu }(p,q)=\chi (k^2)[S_F^{-1}(p)\hat \gamma
_2^a(p,q)-\hat \gamma _1^a(p,q)S_F^{-1}(q)].  \eqnum{9.2}
\end{eqnarray}
From the perturbative calculation, it can be found that In the lowest order,
we have 
\begin{equation}
\begin{array}{c}
\hat \Lambda _\mu ^{(0)a}(p,q)=\gamma _\mu T^a, \\ 
\hat \gamma _1^{(0)a}(p,q)=\hat \gamma _2^{(0)a}(p,q)=T^a.
\end{array}
\eqnum{9.3}
\end{equation}
In the one-loop approximation of order $g^2$, the quark-gluon vertex denoted
by $\hat \Lambda _\mu ^{(1)a}(p,q)$ is contributed from the two diagrams in
Figs. (8a) and (8b) whose expressions can easily be written out. The
quark-ghost vertex functions $\hat \gamma _i^{(1)a}(p,q)$ ($i=1,2$) are
contributed from Figs. (9a) and (9b) and can be represented as 
\begin{equation}
\hat \gamma _i^{(1)a}(p,q)=T^aK_i(p,q)  \eqnum{9.4}
\end{equation}
where 
\begin{equation}
K_1(p,q)=i\frac 32g^2\int \frac{d^4l}{(2\pi )^4}\gamma ^\mu S_F(l)(q-l)^\nu
D_{\mu \nu }(p-l)\Delta (q-l)  \eqnum{9.5}
\end{equation}
and 
\begin{equation}
K_2(p,q)=i\frac 32g^2\int \frac{d^4l}{(2\pi )^4}S_F(l)\gamma ^\mu D_{\mu \nu
}(q-l)(p-l)^\nu \Delta (p-l).  \eqnum{9.6}
\end{equation}
It is clear that the above functions are logarithmically divergent. In the
one-loop approximation, the function $\chi (k^2)$ can be written as $\chi
(k^2)=1-\hat \Omega ^{(1)}(k^2)$ where the ghost particle one-loop
self-energy $\hat \Omega ^{(1)}(k^2)$ was represented in Eq. (8.22). Thus,
up to the order of $g^2$, with setting $\hat \Lambda _\mu
^{(1)a}(p,q)=T^a\Lambda _\mu ^{(1)}(p,q)$, we can write 
\begin{equation}
\hat \Lambda _\mu ^a(p,q)=T^a[\gamma _\mu +\Lambda _\mu ^{(1)}(p,q)] 
\eqnum{9.7}
\end{equation}
and 
\begin{equation}
\chi (k^2)\hat \gamma _i^a(p,q)=T^a[1+I_i(p,q)]  \eqnum{9.8}
\end{equation}
where 
\begin{equation}
I_i(p,q)=K_i(p,q)-\Omega ^{(1)}(k^2).  \eqnum{9.9}
\end{equation}
Upon substituting Eqs. (9.7) and (9.8) and the inverse of the quark
propagator denoted in Eq. (7.17) into Eq. (9.2), then differentiating the
both sides of Eq. (9.2) with respect to $p^\mu $ and finally setting $q=p$,
in the order of $g^2$, we get 
\begin{equation}
\overline{\Lambda }_\mu (p,p)=-\frac{\partial \Sigma (p)}{\partial p^\mu } 
\eqnum{9.10}
\end{equation}
where 
\begin{equation}
\begin{array}{c}
\overline{\Lambda }_\mu (p,p)=\Lambda _\mu ^{(1)}(p,p)-\gamma _\mu I_2(p,p)-(%
{\bf p}-m)\frac{\partial I_2(p,q)}{\partial p^\mu }\mid _{q=p} \\ 
+\frac{\partial I_1(p,q)}{\partial p^\mu }\mid _{q=p}({\bf p}-m)
\end{array}
\eqnum{9.11}
\end{equation}
here $m$ is the $i$-th quark mass (hereafter the subscript of $m_i$ is
suppressed for simplicity). It is emphasized that at one-loop level, the
both sides of Eq. (9.11) are of the order of $g^2$. In the derivation of Eq.
(9.11), the terms of orders higher than $g^2$ have been neglected. The
identity in Eq. (9.10) formally is the same as we met in QED. By the
subtraction at ${\bf p}{=\mu }$, the vertex $\overline{\Lambda }_\mu (p,p)$
will be expressed in the form 
\begin{equation}
\overline{\Lambda }_\mu (p,p)=L\gamma _\mu +\overline{\Lambda }_\mu ^c(p) 
\eqnum{9.12}
\end{equation}
where $L$ is a divergent constant defined by 
\begin{equation}
L=\overline{\Lambda }_\mu (p,p)\mid _{{\bf p}=\mu }  \eqnum{9.13}
\end{equation}
and $\overline{\Lambda }_\mu ^c(p)$ is the finite part of $\overline{\Lambda 
}_\mu (p,p)$ satisfying the boundary condition 
\begin{equation}
\overline{\Lambda }_\mu ^c(p)\mid _{{\bf p}=\mu }=0.  \eqnum{9.14}
\end{equation}
On integrating the identity in Eq. (9.10) over the momentum $p_{\mu \text{ }%
} $and considering the expression in Eq. (9.12), we obtain 
\begin{equation}
\Sigma (p)=A+({\bf p}-\mu )[B-C(p^2)]  \eqnum{9.15}
\end{equation}
where

\begin{equation}
A=\Sigma (\mu ),  \eqnum{9.16}
\end{equation}

\begin{equation}
B=-L  \eqnum{9.17}
\end{equation}
and $C(p^2)$ is defined by 
\begin{equation}
\int_{p_0^\mu }^{p^\mu }dp^\mu \overline{\Lambda }_\mu ^c(p)=({\bf p}-\mu
)C(p^2).  \eqnum{9.18}
\end{equation}
Clearly, the expression in Eq. (9.15) gives the subtraction version for the
quark self-energy which is required by the W-T identity and correct at least
in the approximation of order $g^2$. With this subtraction, the quark
propagator in Eq. (7.17) will be renormalized as 
\begin{equation}
S_F(p)=\frac{Z_2}{{\bf p}-m_R-\Sigma _R(p)}  \eqnum{9.19}
\end{equation}
where $Z_2$ is the renormalization constant defined by 
\begin{equation}
Z_2^{-1}=1-B,  \eqnum{9.20}
\end{equation}
$m_R$ is the renormalized quark mass defined as 
\begin{equation}
m_R=Z_m^{-1}m  \eqnum{9.21}
\end{equation}
in which 
\begin{equation}
Z_m^{-1}=1+Z_2[Am^{-1}+(1-\mu m^{-1})B],  \eqnum{9.22}
\end{equation}
$Z_m$ is the quark mass renormalization constant and $\Sigma _R(p)$ is the
finite correction of the self-energy satisfying the boundary condition $%
\Sigma _R(p)_{\mid p^2=\mu ^2}=0.$

Now we are in a position to discuss the one-loop renormalization of quark
mass. The RGE for the renormalized quark mass can be written from Eq. (8.2)
by setting $F=m$, 
\begin{equation}
\lambda \frac{dm_R(\lambda )}{d\lambda }+\gamma _m(\lambda )m_R(\lambda )=0 
\eqnum{9.23}
\end{equation}
where 
\begin{equation}
\gamma _m(\lambda )=\mu \frac d{d\mu }\ln Z_m.  \eqnum{9.24}
\end{equation}
It is clear that to determine the one-loop renormalization constant $Z_m$,
we first need to determine the divergent constants $A$ and $B$ from the
self-energy represented in Eq. (9.15). The one-loop self-energy denoted by $%
-i\Sigma (p)$ can be written out from Fig. (10). In the Feynman gauge, it is 
\begin{equation}
\Sigma (p)=-i\frac 43g^2\int \frac{d^4k}{(2\pi )^4}\frac{\gamma ^\mu ({\bf k}%
+{\bf p}+m)\gamma _\mu }{[(k+p)^2-m^2+i\varepsilon ](k^2-M^2+i\varepsilon )}
\eqnum{9.25}
\end{equation}
where ${\bf k=}\gamma ^\mu k_\mu $ and ${\bf p=}\gamma ^\mu p_\mu .$ By
making use of the dimensional regularization to calculate the above
integral, it is found that 
\begin{equation}
\Sigma (p)=\frac 83\frac{g^2}{(4\pi )^2}\int_0^1dx\frac{(x-1){\bf p+}2m}{%
\varepsilon [p^2x(x-1)+m^2x+M^2(1-x)]^\varepsilon }.  \eqnum{9.26}
\end{equation}
According to Eq. (9.16), we have 
\begin{equation}
A=\Sigma (p)\mid _{{\bf p}=\mu }=\frac 83\frac{g^2}{(4\pi )^2}\int_0^1dx%
\frac{(x-1)\mu {\bf +}2m}{\varepsilon [\mu
^2x(x-1)+m^2x+M^2(1-x)]^\varepsilon }.  \eqnum{9.27}
\end{equation}
With the aid of the following formula 
\begin{equation}
\frac 1{a^\varepsilon }-\frac 1{b^\varepsilon }=\varepsilon \int_0^1dx\frac{%
b-a}{[ax+b(1-x)]^{1+\varepsilon }},  \eqnum{9.28}
\end{equation}
one can get from Eqs. (9.15) and (9.27) that 
\begin{equation}
\begin{array}{c}
B=[\Sigma (p)-A]({\bf p-}m{\bf )}^{-1}\mid _{{\bf p}=\mu } \\ 
=\frac 83\frac{g^2}{(4\pi )^2}\int_0^1dx\{\frac{(x-1)}{\varepsilon [\mu
^2x(x-1)+m^2x+M^2(1-x)]^\varepsilon }-\frac{2x(x-1)[(x-1)\mu ^2+2m\mu ]}{\mu
^2x(x-1)+m^2x+M^2(1-x)}\}
\end{array}
\eqnum{9.29}
\end{equation}
where $C(\mu ^2)=0$ has been considered. On inserting Eqs. (9.27) and (9.29)
into Eq. (9.22) and noting that in the approximation of order $g^2,$ $%
Z_2\simeq 1$ should be taken in Eq. (9.22), it can be found that 
\begin{equation}
\begin{array}{c}
Z_m=1-\frac Am-(1-\frac \mu m)B \\ 
=1-\frac{g^2}{(4\pi )^2}\frac 83\int_0^1dx\{\frac{(x+1)}{\varepsilon [\mu
^2x(x-1)+m^2x+M^2(1-x)]^\varepsilon }+\frac{2x(x-1)[(x-1)(\mu /m-1)\mu
^2+2(\mu ^2-m\mu ]}{\mu ^2x(x-1)+m^2x+M^2(1-x)}\}.
\end{array}
\eqnum{9.30}
\end{equation}
When substituting Eq. (9.30) in Eq. (9.24) and applying the familiar
integration formulas, through a lengthy calculation, we obtain 
\begin{equation}
\begin{array}{c}
\gamma _m(\lambda )=\frac{4\alpha _R}{3\pi }\{\xi _1(\lambda )+\xi
_2(\lambda )\ln \frac \beta \rho +\frac 2{\lambda ^2K(\lambda )}\xi
_3(\lambda ) \\ 
+\frac 1{2\lambda ^4}[\xi _4(\lambda )+\frac 4{K(\lambda )}\xi _5(\lambda
)]J(\lambda )\}
\end{array}
\eqnum{9.31}
\end{equation}
where 
\begin{equation}
\xi _1(\lambda )=\frac 1{\lambda ^2}[\frac{\lambda ^3}{2\rho }+\frac 32%
\lambda ^2+(\beta ^2-3\rho ^2)\frac \lambda \rho +\rho ^2-\beta ^2], 
\eqnum{9.32}
\end{equation}
\begin{equation}
\begin{array}{c}
\xi _2(\lambda )=\frac 1{\lambda ^4}[\frac{\beta ^2}\rho \lambda ^3+(6\rho
^2-7\beta ^2)\lambda ^2 \\ 
-\frac 3\rho (\rho ^2-\beta ^2)(3\rho ^2-\beta ^2)\lambda +3(\rho ^2-\beta
^2)^2],
\end{array}
\eqnum{9.33}
\end{equation}
\begin{equation}
\begin{array}{c}
\xi _3(\lambda )=\frac{\beta ^2}\rho \lambda ^5-(2\rho ^2+3\beta ^2)\lambda
^4+\frac 1\rho (3\rho ^4+3\rho ^2\beta ^2-2\beta ^4)\lambda ^3 \\ 
+(\rho ^2-\beta ^2)(\rho ^2-4\beta ^2)\lambda ^2-\frac 1\rho (\rho ^2-\beta
^2)(3\rho ^4-4\rho ^2\beta ^2+\beta ^4)\lambda +(\rho ^2-\beta ^2)^3,
\end{array}
\eqnum{9.34}
\end{equation}
\begin{equation}
\begin{array}{c}
\xi _4(\lambda )=\frac{\beta ^2}\rho \lambda ^5-(6\rho ^2+7\beta ^2)\lambda
^4+\frac 1\rho (9\rho ^4+11\rho ^2\beta ^2-8\beta ^4)\lambda ^3 \\ 
+11(\rho ^2-\beta ^2)(\rho ^2-2\beta ^2)\lambda ^2-\frac 7\rho (\rho
^2-\beta ^2)^2(3\rho ^2-\beta ^2)\lambda +7(\rho ^2-\beta ^2)^3,
\end{array}
\eqnum{9.35}
\end{equation}
\begin{equation}
\begin{array}{c}
\xi _5(\lambda )=\frac{\beta ^4}\rho \lambda ^7-(2\rho ^4+3\beta ^4)\lambda
^6+\frac 1\rho (3\rho ^6+4\rho ^2\beta ^4-3\beta ^6)\lambda ^5 \\ 
+(\rho ^2-\beta ^2)(3\rho ^4-\rho ^2\beta ^2-7\beta ^4)\lambda ^4-\frac 1\rho
(\rho ^2-\beta ^2)^2(6\rho ^4+5\rho ^2\beta ^2-3\beta ^4)\lambda ^3 \\ 
+5\beta ^2(\rho ^2-\beta ^2)^3\lambda ^2+\frac 1\rho (\rho ^2-\beta
^2)^4(3\rho ^2-\beta ^2)\lambda -(\rho ^2-\beta ^2)^5,
\end{array}
\eqnum{9.36}
\end{equation}

\begin{equation}
K(\lambda )=\lambda ^4-2(\beta ^2+\rho ^2)\lambda ^2+(\rho ^2-\beta ^2)^2 
\eqnum{9.37}
\end{equation}
and 
\begin{equation}
\begin{array}{c}
\gamma _m(\lambda )=\frac{4\alpha _R}{3\pi }\{\xi _1(\lambda )+\xi
_2(\lambda )\ln \frac \beta \rho +\frac 2{\lambda ^2K(\lambda )}\xi
_3(\lambda ) \\ 
+\frac 1{2\lambda ^4}[\xi _4(\lambda )+\frac 4{K(\lambda )}\xi _5(\lambda
)]J(\lambda )\}
\end{array}
\eqnum{9.31}
\end{equation}
where 
\begin{equation}
\xi _1(\lambda )=\frac 1{\lambda ^2}[\frac{\lambda ^3}{2\rho }+\frac 32%
\lambda ^2+(\beta ^2-3\rho ^2)\frac \lambda \rho +\rho ^2-\beta ^2], 
\eqnum{9.32}
\end{equation}
\begin{equation}
\begin{array}{c}
\xi _2(\lambda )=\frac 1{\lambda ^4}[\frac{\beta ^2}\rho \lambda ^3+(6\rho
^2-7\beta ^2)\lambda ^2 \\ 
-\frac 3\rho (\rho ^2-\beta ^2)(3\rho ^2-\beta ^2)\lambda +3(\rho ^2-\beta
^2)^2],
\end{array}
\eqnum{9.33}
\end{equation}
\begin{equation}
\begin{array}{c}
\xi _3(\lambda )=\frac{\beta ^2}\rho \lambda ^5-(2\rho ^2+3\beta ^2)\lambda
^4+\frac 1\rho (3\rho ^4+3\rho ^2\beta ^2-2\beta ^4)\lambda ^3 \\ 
+(\rho ^2-\beta ^2)(\rho ^2-4\beta ^2)\lambda ^2-\frac 1\rho (\rho ^2-\beta
^2)(3\rho ^4-4\rho ^2\beta ^2+\beta ^4)\lambda +(\rho ^2-\beta ^2)^3,
\end{array}
\eqnum{9.34}
\end{equation}
\begin{equation}
\begin{array}{c}
\xi _4(\lambda )=\frac{\beta ^2}\rho \lambda ^5-(6\rho ^2+7\beta ^2)\lambda
^4+\frac 1\rho (9\rho ^4+11\rho ^2\beta ^2-8\beta ^4)\lambda ^3 \\ 
+11(\rho ^2-\beta ^2)(\rho ^2-2\beta ^2)\lambda ^2-\frac 7\rho (\rho
^2-\beta ^2)^2(3\rho ^2-\beta ^2)\lambda +7(\rho ^2-\beta ^2)^3,
\end{array}
\eqnum{9.35}
\end{equation}
\begin{equation}
\begin{array}{c}
\xi _5(\lambda )=\frac{\beta ^4}\rho \lambda ^7-(2\rho ^4+3\beta ^4)\lambda
^6+\frac 1\rho (3\rho ^6+4\rho ^2\beta ^4-3\beta ^6)\lambda ^5 \\ 
+(\rho ^2-\beta ^2)(3\rho ^4-\rho ^2\beta ^2-7\beta ^4)\lambda ^4-\frac 1\rho
(\rho ^2-\beta ^2)^2(6\rho ^4+5\rho ^2\beta ^2-3\beta ^4)\lambda ^3 \\ 
+5\beta ^2(\rho ^2-\beta ^2)^3\lambda ^2+\frac 1\rho (\rho ^2-\beta
^2)^4(3\rho ^2-\beta ^2)\lambda -(\rho ^2-\beta ^2)^5,
\end{array}
\eqnum{9.36}
\end{equation}

\begin{equation}
K(\lambda )=\lambda ^4-2(\beta ^2+\rho ^2)\lambda ^2+(\rho ^2-\beta ^2)^2 
\eqnum{9.37}
\end{equation}
and 
\begin{equation}
\begin{array}{c}
\gamma _m(\lambda )=\frac{4\alpha _R}{3\pi }\{\xi _1(\lambda )+\xi
_2(\lambda )\ln \frac \beta \rho +\frac 2{\lambda ^2K(\lambda )}\xi
_3(\lambda ) \\ 
+\frac 1{2\lambda ^4}[\xi _4(\lambda )+\frac 4{K(\lambda )}\xi _5(\lambda
)]J(\lambda )\}
\end{array}
\eqnum{9.31}
\end{equation}
where 
\begin{equation}
\xi _1(\lambda )=\frac 1{\lambda ^2}[\frac{\lambda ^3}{2\rho }+\frac 32%
\lambda ^2+(\beta ^2-3\rho ^2)\frac \lambda \rho +\rho ^2-\beta ^2], 
\eqnum{9.32}
\end{equation}
\begin{equation}
\begin{array}{c}
\xi _2(\lambda )=\frac 1{\lambda ^4}[\frac{\beta ^2}\rho \lambda ^3+(6\rho
^2-7\beta ^2)\lambda ^2 \\ 
-\frac 3\rho (\rho ^2-\beta ^2)(3\rho ^2-\beta ^2)\lambda +3(\rho ^2-\beta
^2)^2],
\end{array}
\eqnum{9.33}
\end{equation}
\begin{equation}
\begin{array}{c}
\xi _3(\lambda )=\frac{\beta ^2}\rho \lambda ^5-(2\rho ^2+3\beta ^2)\lambda
^4+\frac 1\rho (3\rho ^4+3\rho ^2\beta ^2-2\beta ^4)\lambda ^3 \\ 
+(\rho ^2-\beta ^2)(\rho ^2-4\beta ^2)\lambda ^2-\frac 1\rho (\rho ^2-\beta
^2)(3\rho ^4-4\rho ^2\beta ^2+\beta ^4)\lambda +(\rho ^2-\beta ^2)^3,
\end{array}
\eqnum{9.34}
\end{equation}
\begin{equation}
\begin{array}{c}
\xi _4(\lambda )=\frac{\beta ^2}\rho \lambda ^5-(6\rho ^2+7\beta ^2)\lambda
^4+\frac 1\rho (9\rho ^4+11\rho ^2\beta ^2-8\beta ^4)\lambda ^3 \\ 
+11(\rho ^2-\beta ^2)(\rho ^2-2\beta ^2)\lambda ^2-\frac 7\rho (\rho
^2-\beta ^2)^2(3\rho ^2-\beta ^2)\lambda +7(\rho ^2-\beta ^2)^3,
\end{array}
\eqnum{9.35}
\end{equation}
\begin{equation}
\begin{array}{c}
\xi _5(\lambda )=\frac{\beta ^4}\rho \lambda ^7-(2\rho ^4+3\beta ^4)\lambda
^6+\frac 1\rho (3\rho ^6+4\rho ^2\beta ^4-3\beta ^6)\lambda ^5 \\ 
+(\rho ^2-\beta ^2)(3\rho ^4-\rho ^2\beta ^2-7\beta ^4)\lambda ^4-\frac 1\rho
(\rho ^2-\beta ^2)^2(6\rho ^4+5\rho ^2\beta ^2-3\beta ^4)\lambda ^3 \\ 
+5\beta ^2(\rho ^2-\beta ^2)^3\lambda ^2+\frac 1\rho (\rho ^2-\beta
^2)^4(3\rho ^2-\beta ^2)\lambda -(\rho ^2-\beta ^2)^5,
\end{array}
\eqnum{9.36}
\end{equation}

\begin{equation}
K(\lambda )=\lambda ^4-2(\beta ^2+\rho ^2)\lambda ^2+(\rho ^2-\beta ^2)^2 
\eqnum{9.37}
\end{equation}
and 
\begin{equation}
J(\lambda )=\frac 1{\sqrt{K(\lambda )}}\ln \frac{\lambda ^2-(\beta ^2+\rho
^2)-\sqrt{K(\lambda )}}{\lambda ^2-(\beta ^2+\rho ^2)+\sqrt{K(\lambda )}}. 
\eqnum{9.38}
\end{equation}

With the anomalous dimension given above, the equation in Eq. (9.23) can be
solved and gives the effective mass for a quark as follows 
\begin{equation}
m_R(\lambda )=m_Re^{-S_q(\lambda )}  \eqnum{9.39}
\end{equation}
where 
\begin{equation}
S_q(\lambda )=\int_1^\lambda \frac{d\lambda }\lambda \gamma _m(\lambda ). 
\eqnum{9.40}
\end{equation}
This integral is not able to be analytically calculated even though the
coupling constant in Eq. (9.31) is taken to be a constant. In the large
momentum limit ($\lambda \rightarrow \infty $), Eq. (9.31) tends to 
\begin{equation}
\gamma _m(\lambda )\approx \frac{2\alpha _R}{3\pi \rho }\lambda . 
\eqnum{9.41}
\end{equation}
In this limit, it is seen that in the time-like momentum space, we have 
\begin{equation}
m_R(\lambda )\approx m_Re^{-\frac{2\alpha _R}{3\pi \rho }\lambda
}\rightarrow 0.  \eqnum{9.42}
\end{equation}

Graphically, we only show in Fig. (11) the $s$ quark effective mass $%
m_R(\lambda )$ given by the timelike momentum subtraction with the coupling
constant being taken to be a constant. For other quarks, the behavior of
their effective masses is similar. In Fig. (11), the solid line represents
the effective mass given by the massive QCD, while the dashed line
represents the effective mass given by the massless QCD. From the figure, we
see that the effective mass with a finite gluon mass behaves as a constant
equal to the mass $m_R$ in the region [0,1] of $\lambda $ and then rapidly
falls to zero when $\lambda $ goes from unit to infinity. The effective mass
with zero-gluon mass has a peak around $\lambda =1.38$. The appreciable
difference between the both effective masses occurs in the region [0,10] of $%
\lambda $. The effective quark mass given by the spacelike momentum
subtraction can directly be written out from Eqs. (9.31)-(9.40) by replacing
the $\lambda $ in $\gamma _m(\lambda )$ with $i\lambda $ and hence the $%
m_R(\lambda )$ becomes a complex function. By numerical calculations, it is
found that either the real part or the imaginary part of the $m_R(\lambda )$
behaves as an oscillating function with a damping amplitude as exhibited in
Fig. (12). In the figure, the solid and the dashed lines represent
respectively the real part and imaginary part of the effective quark mass
which is given by the massive QCD, while the dotted line shows the real part
of the effective quark mass which is given by the massless QCD. It is noted
that in the most of practical applications to the both of scattering and
bound state problems, only the effective quark mass given by the timelike
momentum subtraction is concerned .

\section{Conclusions and discussions}

In this paper, it has been shown that as the massless QCD, the massive QCD
established on the basis of gauge-invariance has a set of
BRST-transformations under which the effective action and generating
functional are invariant. From the BRST-invariance, we derived a set of W-T
identities satisfied by the generating functionals for full Green functions,
connected Green functions and proper vertex functions. Furthermore, from the
above identities, we derived the W-T identities respected by the gluon
propagator, the three-line and four-line proper gluon vertices and the
quark-gluon proper vertex. Based on these identities we discussed the
renormalization of the propagators and the vertices. In particular, from the
renormalized forms of the W-T identities obeyed by propagators and vertices,
the S-T identity for the renormalization constants is naturally deduced.
This identity is helpful for the renormalization by means of the
renormalization group approach. To show the renormalizability of the massive
QCD, the one-loop renormalization is performed by the renormalization group
method. In this renormalization, the analytical expressions of the one-loop
effective coupling constant, gluon mass and quark mass have been derived.
Since the renormalization was carried out by employing the mass-dependent
momentum space subtraction scheme and exactly respecting the W-T identities,
the results obtained are faithful and allow us to discuss the physical
behaviors of the effective coupling constant and masses in the whole range
of momentum (or distance). Particularly, the previous result given in the
minimal subtraction scheme for massless QCD is naturally recovered in the
large momentum limit.

As shown in sections 8 and 9, in the mass-dependent renormalization, it is
necessary to distinguish the results given by the timelike momentum
subtraction from the corresponding ones obtained by the spacelike momentum
subtraction. For example, one can see from Figs. (4) and (5) that the
effective coupling constants given in the timelike and spacelike subtraction
schemes have different behaviors in the low and intermediate energy region
although in the large momentum limit, the difference between the both
coupling constants disappears. Obviously, the both results obtained in the
timelike and spacelike momentum subtraction schemes are meaningful and
suitable for different physical processes. For instance, when we study the
quark-quark scattering taking place in the t-channel, the transfer momentum
in the gluon propagator is spacelike. In this case, it is suitable to take
the effective coupling constant and gluon mass given by the spacelike
momentum subtraction. If we investigate the quark-antiquark annihilation
process which takes place in the s-channel, since the transfer momentum is
timelike, the effective coupling constant and gluon mass given in the
timelike momentum subtraction scheme should be used. It is also seen from
sections 8 and 9 that the gluon mass gives a considerable effect on the
behaviors of the effective coupling constant and particle masses. In
particular, the gluon mass plays an crucial role to determine the singular
or analytical behavior of the effective coupling constant. Since the
gauge-invariance does not exclude the gluon to have a mass, it is
interesting to examine the gluon mass effect on physical processes. At
present, the massless QCD has widely been recognized to be the candidate of
the strong interaction theory and has been proved to be compatible with the
present high energy experiments. However, we think, the massive QCD would be
more favorable to explain the strong interaction phenomenon, particularly,
at the low energy region because the massive gluon would make the force
range more shorter than that caused by the massless gluon. As for the high
energy and large momentum transfer phenomena, as seen from the massive gluon
propagator, the gluon mass gives little influence on the theoretical result
in this case so that the massive QCD could not conflict with the
well-established results gained from the massless QCD in the high energy
domain.

At last, we would like to make some remarks on the nilpotency problem of the
BRST-external sources. In comparison of the BRST-transformations and the W-T
identities for the massive QCD with those for the massless QCD, it is seen
that they formally are almost the same. The only difference is that the
BRST-transformation for the ghost particle field $C^a(x)$ written in Eq.
(2.25) has an extra term proportional to $\sigma ^2$, the ghost particle
mass squared. Due to this term, there also appear some $\sigma ^2-$dependent
terms in the W-T identities as denoted in Eqs. (3.18), (4.2), (4.8), (5.11)
and (6.18). But, in the physical Landau gauge where $\sigma =0$, all the $%
\sigma ^2$-dependent terms disappear. In this case, all the
BRST-transformations and W-T identities for the massive QCD are formally
identical to those for the massless QCD. As one knows, for the massless QCD,
the composite field functions $\Delta \Phi _i$ as defined in Eq. (3.3) have
the nilpotency property: $\delta \Delta \Phi _i=0$ under the
BRST-transformations [13-16] which guarantee the BRST-invariance of the
BRST-source terms introduced in the generating functional. This nilpotency
property is still preserved for the massive QCD established in the Landau
gauge because in the Landau gauge, the BRST-transformations are identical to
those for the massless QCD. However, for the massive QCD set up in arbitrary
gauges, we find $\delta \Delta \Phi _i$ $\neq 0$, the nilpotency loses due
to nonzero of the ghost particle spurious mass $\sigma $. In this case, as
pointed out in section 3, to ensure the BRST-invariance of the source terms,
we may simply require the sources $u_i$ to satisfy the condition denoted in
Eq. (3.7). The definition in Eq. (3.7) for the sources is reasonable. Why
say so? Firstly, we note that the original W-T identity formulated in Eq.
(3.2) does not involve the BRST- sources. This identity is suitable to use
in practical applications. Introduction of the BRST source terms in the
generating functional is only for the purpose of representing the identity
in Eq. (3.2) in a convenient form, namely, to represent the composite field
functions in the identity in terms of the differentials of the generating
functional with respect to the corresponding sources. For this purpose, we
may start from the generating functional defined in Eq. (3.4) to re-derive
the identity in Eq. (3.2). In doing this, it is necessary to require the
source terms $u_i\triangle \Phi _i$ to be BRST-invariant so as to make the
derived identity coincide with that given in Eq. (3.2). How to ensure the
source terms to be BRST-invariant? If the composite field functions $%
\triangle \Phi _i$ are nilpotent under the BRST-transformation, the
BRST-invariance of the source terms is certainly guaranteed. Nevertheless,
the nilpotency of the functions $\triangle \Phi _i$ is not a uniquely
necessary condition to ensure the BRST- invariance of the source terms,
particularly, in the case where the functions $\triangle \Phi _i$ are not
nilpotent. In the latter case, considering that under the BRST-
transformations, the functions $\triangle \Phi _i$ can be, in general,
expressed as $\delta \Delta \Phi _i=\xi \tilde \Phi _i$ where the $\tilde 
\Phi _i$ are some nonvanishing functions, we may alternatively require the
sources $u_i$ to satisfy the condition shown in Eq. (3.7) so as to guarantee
the source terms to be BRST- invariant. Actually, this is a general trick to
make the source terms to be BRST-invariant in spite of whether the functions 
$\triangle \Phi _i$ are nilpotent or not. As mentioned before, the sources
themselves have no physical meaning. They are, as a mathematical tool,
introduced into the generating functional just for performing the
differentiations. For this purpose, only a certain algebraic and analytical
properties of the sources are necessarily required. Particularly, in the
differentiations, only the infinitesimal property of the sources are
concerned. Therefore, the sources defined in Eq. (3.7) are mathematically
suitable for the purpose of introducing them. The reasonability of the
arguments stated above for the source terms is substantiated by the
correctness of the W-T identities derived in sections 4-7. Even though the
identities in Eqs. (4.1) and (4.2) are derived from the W-T identity in Eq.
(3.8) which is represented in terms of the differentials with respect to the
BRST-sources, they give rise to correct relations between the propagators
and/or vertices. For example, the correctness of the relation in Eq. (4.8)
can easily be verified by the free propagators written in Eqs. (4.10) and (4
14). These propagators are usually derived from the generating functional in
Eq. (2.12) by employing the perturbation method without concerning the
BRST-source terms and the nilpotency of the BRST- transformations. A
powerful argument of proving the correctness of the way of introducing the
BRST-sources is that after completing the differentiations in Eq. (3.8) and
setting the BRST-sources to vanish, we immediately obtain the W-T identity
in Eq. (3.2) which is irrelevant to the BRST-sources. Therefore, all
identities or relations derived from the W-T identity in Eq. (3.8) are
completely the same as those derived from the identity in Eq. (3.2). An
important example of showing this point will be presented in Appendix where
an identity derived from the W-T identities in Eqs. (3.8) can equally be
derived from the generating functional in Eq. (2.12) which does not involve
the BRST-sources.

\section{Acknowledgment}

This work was supported by National Natural Science Foundation of China.

\section{Appendix: Alternative derivation of the W-T identity}

In this appendix, we give an alternative derivation of a W-T identity
without concerning the nilpotency of the composite field functions appearing
in the BRST-source terms and show that the result is equal to the one
obtained from . generating functional containing the BRST-external sources.
Let us start from the generating functional of Green functions given in Eqs.
(2.12) and (2.13). For simplicity of statement, we omit the fermion field
functions in the generating functional and rewrite the functional in the
form 
\begin{equation}
\begin{array}{c}
Z[J,\overline{\xi },\xi ]=\frac 1N\int {\cal D}[A.\bar C.C]exp\{iS+i\int
d^4x[-\frac 1{2\alpha }(\partial ^\mu A_\mu ^a)^2+J^{a\mu }A_\mu ^a \\ 
+\overline{\xi }^aC^a+\bar C^a\xi ^a]+i\int d^4xd^4y\bar C%
^a(x)M^{ab}(x,y)C^b(y)\}
\end{array}
\eqnum{A1}
\end{equation}
where 
\begin{equation}
S=\int d^4x[-\frac 14F^{a\mu \nu }F_{\mu \nu }^a+\frac 12M^2A^{a\mu }A_\mu
^a]  \eqnum{A2}
\end{equation}
and 
\begin{equation}
M^{ab}(x,y)=\partial _x^\mu [{\cal D}_\mu ^{ab}(x)\delta ^4(x-y)]  \eqnum{A3}
\end{equation}
in which ${\cal D}_\mu ^{ab}(x)$ was defined in Eq. (2.9).

When we make the following translation transformations in Eq. (A1) 
\begin{equation}
\begin{array}{c}
C^a(x)\to C^a(x)-\int d^4y(M^{-1})^{ab}(x,y)K^b(y) \\ 
\bar C^a(x)\to \bar C^a(x)-\int d^4y\bar K^b(y)(M^{-1})^{ba}(y,x)
\end{array}
\eqnum{A4}
\end{equation}
and complete the integration over the ghost field variables, Eq. (A1) will
be expressed as 
\begin{equation}
Z[J,\bar K,K]=e^{-i\int d^4xd^4y\overline{\xi }^a(x)(M^{-1})^{ab}(x,y,\delta
/i\delta J)\xi ^b(y)}Z[J]  \eqnum{A5}
\end{equation}
where $Z[J]$ is the generating functional without the external sources of
ghost fields [13,17] 
\begin{equation}
Z[J]=\frac 1N\int {\cal D}(A)\Delta _F[A]exp\{iS+i\int d^4x[-\frac 1{2\alpha 
}(\partial ^\mu A_\mu ^a)^2+J^{a\mu }A_\mu ^a]\}  \eqnum{A6}
\end{equation}
in which 
\begin{equation}
\Delta _F[A]=detM[A]  \eqnum{A7}
\end{equation}
here the matrix $M[A]$ was defined in Eq. (A3). From Eq. (A5), we may obtain
the ghost particle propagator in the presence of the external source $J$ 
\begin{equation}
\begin{array}{c}
i\Delta ^{ab}[x,y,J]=\frac{\delta ^2Z[J,\bar K,K]}{\delta \overline{\xi }%
^a(x)\delta \xi ^b(y)}|_{\overline{K}=K=0} \\ 
=i(M^{-1})_{ab}[x,y,\frac \delta {i\delta J}]Z[J].
\end{array}
\eqnum{A8}
\end{equation}
The above result allows us to rewrite the W-T identity in Eq. (4.1) in terms
of the generating functional $Z[J]$ when completing the derivative with
respect to $u^{b\nu }(y)$ and setting $\overline{\xi }^a(x)=\xi ^b(y)=0$, 
\begin{equation}
\frac 1\alpha \partial _x^\mu \frac{\delta Z[J]}{i\delta J^{a\mu }(x)}-\int
d^4yJ^{b\mu }(y)D_\mu ^{bd}[y,\frac \delta {i\delta J}](M^{-1})^{da}(y,x,%
\frac \delta {i\delta J})Z[J]=0  \eqnum{A9}
\end{equation}
where 
\begin{equation}
D_\mu ^{bd}(y)={\cal D}_\mu ^{bd}(y)-\frac{\sigma ^2}{\Box _y}\partial _\mu
^y\delta ^{bd}  \eqnum{A10}
\end{equation}
is the ordinary covariant derivative. On completing the differentiations
with respect to the source $J$, Eq. (A9) reads 
\begin{equation}
\begin{array}{c}
\frac 1N\int {\cal D}[A]\Delta _F[A]exp\{iS+i\int d^4x[-\frac 1{2\alpha }%
(\partial ^\mu A_\mu ^a)^2+J^{a\mu }A_\mu ^a]\} \\ 
\times [\int d^4yJ^{b\mu }(y)D_\mu ^{bc}(y)(M^{-1})^{ca}(y,x)-\frac 1\alpha
\partial ^\nu A_\nu ^a(x)] \\ 
=0
\end{array}
\eqnum{A11}
\end{equation}

By making use of Eqs. (A3), (A8) and (A10), the ghost equation shown in Eq.
(4.2) may be written as 
\begin{equation}
M^{ac}[x,\frac \delta {i\delta J}](M^{-1})^{cb}[x,y,\frac \delta {i\delta J}%
]Z[J]=\delta ^{ab}\delta ^4(x-y)Z[J]  \eqnum{A12}
\end{equation}
When the source $J$ is turned off, we get the equation written in Eq. (2.24)
which affirms the fact that the ghost particle propagator is just the
inverse of the matrix $M$.

To confirm the correctness of the identity given in Eq. (A11), we derive the
identity newly by starting from the generating functional written in Eq.
(A6) which does not involve the BRST-external sources. Let us make the
ordinary gauge transformation $\delta A_\mu ^a=D_\mu ^{ab}\theta ^b$ to the
generating functional in Eq. (A6). Considering the gauge-invariance of the
functional integral, the integration measure and the functional $\triangle
_F[A]=\det M[A],$ we get [13,17] 
\begin{equation}
\begin{array}{c}
\delta Z[J]=\frac 1N\int D(A)\triangle _F[A]\int d^4y[J^{b\mu
}(y)+M^2A^{b\mu }(y) \\ 
-\frac 1\alpha \partial ^\nu A_\nu ^b\partial _y^\mu ]D_\mu ^{bc}(y)\theta
^c(y)\exp \{iS+i\int d^4x[-\frac 1{2\alpha }(\partial ^\mu A_\mu
^a)^2+J^{a\mu }A_\mu ^a]\} \\ 
=0
\end{array}
\eqnum{A13}
\end{equation}
According to the well-known procedure, the group parameter $\theta ^a(x)$ in
Eq. (A13) may be determined by the following equation [10,14,17] 
\begin{equation}
M^{ab}(x)\theta ^b(x)\equiv \partial _x^\mu ({\cal D}_\mu ^{ab}(x)\theta
^b(x))=\lambda ^a(x)  \eqnum{A14}
\end{equation}
where $\lambda ^a(x)$ is an arbitrary function. When setting $\lambda
^a(x)=0,$ Eq. (A14) will be reduced to the constraint condition on the gauge
group (the ghost equation) which is used to determine the $\theta ^a(x)$ as
a functional of the vector potential $A_\mu ^a(x)$. However, when the
constraint condition is incorporated into the action by the Lagrange
undetermined multiplier method to give the ghost term in the generating
functional, the $\theta ^a(x)$ should be treated as arbitrary function
according to the spirit of Lagrange multiplier method. That is why we may
use Eq. (A16) to determine the functions $\theta ^a(x)$ in terms of the
function $\lambda ^a(x)$ . From Eq. (A14), we solve 
\begin{equation}
\theta ^a(x)=\int d^4x(M^{-1})^{ab}(x-y)\lambda ^b(y)  \eqnum{A15}
\end{equation}
Upon substituting the above expression into Eq. (A13) and then taking
derivative of Eq. (A13) with respect to $\lambda ^a(x),$ we obtain 
\begin{equation}
\begin{array}{c}
\frac 1N\int D(A)\triangle _F[A]\int d^4y[J^{b\mu }(y)+M^2A^{b\mu }(y) \\ 
-\frac 1\alpha \partial _y^\nu A_\nu ^b(y)\partial _y^\mu ]D_\mu
^{bc}(y)(M^{-1})^{ca}(y-x)\exp \{iS+ \\ 
i\int d^4x[-\frac 1{2\alpha }(\partial ^\mu A_\mu ^a)^2+J^{a\mu }A_\mu ^a]\}
\\ 
=0
\end{array}
\eqnum{A16}
\end{equation}
According to the expression denoted in Eq. (2.4) and the identity $%
f^{bcd}A^{c\mu }A_\mu ^d=0$, it is easy to see 
\begin{equation}
A^{b\mu }(y)D_\mu ^{bc}(y)(M^{-1})^{ca}(y-x)=A^{b\mu }(y)\partial _\mu
^y(M^{-1})^{ba}(y-x)  \eqnum{A17}
\end{equation}
By making use of the relation in Eq. (A10), the definition in Eq. (A3) and
the equation in Eq. (A12), we deduce 
\begin{equation}
\begin{array}{c}
\frac 1\alpha \partial _y^\nu A_\nu ^b(y)\partial _y^\mu D_\mu
^{bc}(y)(M^{-1})^{ca}(y-x) \\ 
=\frac 1\alpha \partial ^\nu A_\nu ^b(y)\delta ^4(x-y)-M^2\partial _y^\nu
A_\nu ^b(y)(M^{-1})^{ba}(y-x)
\end{array}
\eqnum{A18}
\end{equation}
On inserting Eqs. (A17) and (A18) into Eq. (A16), we obtain an identity
which is exactly identical to that given in Eq. (A11) although in the above
derivation, we started from the generating functional without containing the
ghost field functions and the BRST-sources and , therefore, the derivation
does not concern the nilpotency of the composite field functions appearing
in the BRST-source terms. This fact indicates that the W-T identities
derived in section 3 are correct and hence the procedure of introducing the
BRST-invariant source terms into the generating functional is completely
reasonable.

\section{References}

[1] J. M. Cornwall and A. Soni, Phys. Lett. {\bf B 120}, 431 (1983).

[2] W. S. Hou and G. G. Wong, Phys. Rev. {\bf D 67}, 034003 (2003).

[3] M. H. Thoma, M. L\"ust and H. J. Mang, J. Phys. G: Nucl. Part. Phys. 
{\bf 18}, 1125 (1992).

[4] J. Y. Cui, J. M. Wu, H. Y. Jin, Phys. Lett. {\bf B 424}, 381 (1998).

[5] J. M. Cornwall and A. Soni, Phys. Lett. {\bf B 120}, 431 (1983).

[6] J. C. Su, IL Nuovo Cimento {\bf 117 B} (2002) 203-218.

[7] J. C. Su and J. X. Chen, Phys. Rev.{\bf \ D 69}, 076002 (2004).

[8] J. C. Su, Proceedings of Institute of Mathematics of NAS of Ukraine,
Vol. {\bf 50}, Part 2, 965 (2004).

[9] L. D. Faddeev and V. N. Popov, Phys. Lett. {\bf B 25}, 29 (1967).

[10] C. Becchi, A. Rouet and R. Stora, Phys. Lett. {\bf B 52} (1974) 344;

Commun. Math. Phys. {\bf 42} (1975) 127; I. V. Tyutin, Lebedev Preprint {\bf %
39} (1975).

[11] J. C. Ward, Phys. Rev. {\bf 77} (1950) 2931.

[12] Y. Takakashi, Nuovo Cimento {\bf 6} (1957) 370.

[13] E. S. Abers and B. W. Lee, Phys. Rep. {\bf C 9} (1973) 1.

[14] B. W. Lee, in Methods in Field Theory (1975), ed. R .Balian and J.
Zinn-Justin.

[15] W. Marciano and H. Pagels, Phys.Rep {\bf 36} (1978) 137.

[16] C. Itzykson and F-B. Zuber, Quantum Field Theory, McGraw-Hill, New York
(1980).

\begin{itemize}
\item[17]  L. D. Faddeev and A. A. Slavnov, Gauge Fields: Introduction to
Quantum Theory, The Benjamin Commings Publishing Company Inc. (1980).

[18] J. C. Su, hep-th/9805192; hep-th/9805193; hep-th/9805194.

[19] A. Slavnov, Theor. and Math. Phys. {\bf 10} (1972) 99, (English
translation).
\end{itemize}

[20] J. C. Taylor, Nucl. Phys. {\bf B 33} (1971) 436.

[21] C. G. Callan, Phys. Rev. {\bf D 2}, 1541 (1970); K. Symanzik, Commun,
Math. Phys. {\bf 18}, 227 (1970).

[22] S. Weinberg, Phys. Rev. {\bf D }8, 3497 (1973).

[23] J. C. Collins and A. J. Macfarlane, Phys. Rev. {\bf D 10}, 1201 (1974)

[24] H. Umezawa and S. Kamefuchi, Nucl. Phys.{\bf \ 23}, 399 (1961).

[25] A. Salam, Phys. Rev. {\bf 127}, 331 (1962).

[26] D .G. Boulware, Ann. Phys. {\bf 56}, 140 (1970).

[27] A. Burnel, Phys. Rev. {\bf D 33, }2981 (1986); {\bf D 33}, 2985 (1986).

[28] R. Delbourgo, S. Twisk and G. Thompson, Int. J. Mod. Phys. {\bf A 3},
435 (1988).

[29] F. J. Dyson, Phys. Rev. {\bf 75} (1949) 1736; J. Schwinger, Proc. Nat.
Acad. Sci. {\bf 37} (1951) 452.

[30] J. C. Su, X. X. Yi and Y.H. Cao, J. Phys. G: Nucl. Part. Phys. {\bf 25}%
, 2325 (1999).

[31] J. C. Su, L. Shan and Y. H. Cao, Commun. Theor. Phys. {\bf 36}, 665
(2000).

[32] J. C. Su and Hai-Jun Wang, Phys. Rev. {\bf C 70}, 044003 (2004).

[33] G. 't Hooft, Nucl. Phys. {\bf B} {\bf 61}, 455 (1973).

[34] W. A. Bardeen, A. J. Buras, D. W. Duke and T. Muta, Phys. Rev. {\bf D} 
{\bf 18}, 3998 (1978); W. A. Bardeen and R. A. J. Buras, Phys. Rev. {\bf D} 
{\bf 20}, 166 (1979).

[35] W. Celmaster and R. J. Gonsalves, Phys. Rev. Lett. {\bf 42}, 1435
(1979); Phys. Rev. {\bf D} {\bf 20}, 1420 (1979);

W. Celmaster and D. Sivers, Phys. Rev. {\bf D} {\bf 23}, 227 (1981).

[36] E. Braaten and J. P. Leveille, Phys. Rev. {\bf D} {\bf 24}, 1369 (1981).

[37] S. N. Gupta and S. F. Radford, Phys. Rev. {\bf D} {\bf 25}, 2690
(1982); J. C. Collins and A. J. Macfarlane, Phys. Rev. {\bf D} {\bf 10},
1201 (1974).

[38] H. D. Politzer. Phys. Rev. Lett. {\bf 30}, 1346 (1973).

[39] D. J. Gross and F. Wilczek, Phys. Rev. Lett. {\bf 30}, 1343 (1973);
Phys. Rev.{\bf \ D} {\bf 8}, 3633 (1973).

[40] H. Georgi and H. D. Politzer, Phys. Rev. {\bf D 14}, 1829 (1976).

\section{Figure captions}

Fig. (1): The one-loop gluon self-energy. The solid, wavy and dashed lines
represent the free quark, gluon and ghost particle propagators respectively.

Fig. (2): The one-loop ghost particle self-energy. The lines represent the
same as in Fig. (1).

Fig. (3): The one-loop ghost-gluon vertices . The lines mark the same as in
Fig. (1).

Fig. (4): The one-loop effective coupling constants ${\alpha _R(\lambda )}$
given by the timelike momentum space subtraction. The solid and dashed lines
represent the coupling constants given by the massive QCD and the massless
QCD, respectively. The dotted line denotes the coupling constant given in {%
the minimal subtraction scheme}.

Fig. (5): The one-loop effective coupling constants ${\alpha _R(\lambda )}$
given by the spacelike momentum space subtraction. The solid and dashed
lines represent the coupling constants given by the massive QCD and the
massless QCD, respectively. The dotted line denotes the coupling constant
given in {the minimal subtraction scheme}. The dashed-dotted line represents
the one-loop effective coupling constants ${\alpha _R(\lambda )}$ given by
the spacelike momentum subtraction for which the gluon mass is taken to be a
smaller value.

Fig. (6): The one-loop effective coupling constants ${\alpha _R(\lambda )}$
given by the timelike momentum space subtraction. The solid, dashed and
dotted lines represent the coupling constants given by taking the quark
flavor $N_f=2,3$ and $4$ respectively.

Fig. (7): The one-loop effective gluon masses $M_R(\lambda )$. The solid and
the dashed lines represent the effective masses given in the timelike and
spacelike subtractions respectively.

Fig. (8): The one-loop quark-gluon vertices. The lines represent the same as
in Fig. (1).

Fig. (9): The one-loop quark-ghost particle vertices. The lines represent
the same as in Fig. (1).

Fig. (10): The one-loop quark self-energy. The lines represent the same as
in Fig. (1).

Fig. (11): The one-loop effective quark masses $m_R(\lambda )$ given by the
timelike momentum space subtraction. The solid and the dashed lines
represent the effective masses given by the massive QCD and massless QCD,
respectively.

Fig. (12): The one-loop effective quark masses $m_R(\lambda )$ given by the
spacelike momentum space subtraction. The solid and the dashed lines
represent respectively the real part and imaginary part of the effective
quark mass which is given by the massive QCD. The dotted line shows the real
part of the effective quark mass which is given by the massless QCD.

\end{document}